\documentclass[aps,twocolumn,
preprintnumbers,showpacs,showkeys]{revtex4-1}
\usepackage{epsfig,dsfont}
\usepackage{amsmath}
\begin{document}  

\preprint{HU-EP-15/35}

\title{Dual simulation of the 2-dimensional lattice U(1) gauge-Higgs model \\
with a topological term}
\author{Christof Gattringer$\,^a$}
\author{Thomas Kloiber$\,^a$}
\author{Michael M\"uller-Preussker$\,^b$}
\affiliation{$^a$Universit\"at Graz, Institut f\"ur Physik, Universit\"atsplatz 5, 8010 Graz, Austria}
\affiliation{$^b$Humboldt-Universit\"at zu Berlin, Institut f\"ur Physik, 12489 Berlin, Germany}

\begin{abstract}
The 2-dimensional U(1) gauge-Higgs model with a topological term is a simple example of a lattice field theory where the
complex action problem comes from the topological term. We show that the model can be exactly rewritten in terms of dual 
variables, such that the dual partition sum has only real and positive contributions. Using suitable algorithms the dual formulation 
allows for Monte Carlo simulations at arbitrary values of the vacuum angle. We demonstrate the feasibility of the dual simulation
and study the continuum limit, as well as the phase diagram of the system. 
\vskip5mm
{\sl While working on the reply to the referee report the sad news reached us that our friend and co-author 
Michael M\"uller-Preussker has passed away on 12th of October 2015. Michael has devoted 50 years of his 
life to physics as a very successful researcher, but also was a colleague for whom service to the community 
was an important duty. Michael was an inspiration for many and his untimely early death is a big loss for all of us.}
\end{abstract}

\date{July 31, 2015}

\maketitle

\vskip20mm

\section{Introductory remarks}

Systems with a vacuum term have a complex action problem that is similar to the one for systems with finite chemical potential. For the latter 
case it was recently shown that some lattice models can be rewritten exactly to new degrees of freedom, so-called dual variables.
In terms of the dual variables, which are loops for matter fields and surfaces for gauge fields, the partition function has only real and 
positive contributions and Monte Carlo simulations are possible. For examples relevant to this project see 
\cite{Gattringer:2012df,Gattringer:2012ap,Mercado:2013ola,Mercado:2013yta,Kloiber:2014qea,Gattringer:2015nea,Gattringer:2014nxa}.

For clarity we stress that in the strict sense the variant of duality applied here is only the first half of the conventional duality transformation, as will 
also become clear below. On conventional duality transformations there exists extensive literature -- see, e.g., the review \cite{Savit:1979ny}. We remark,
that the second step of the conventional duality transformation, i.e., satisfying the constraints by introducing variables on the dual lattice is often not 
possible for the models considered in 
\cite{Gattringer:2012df,Gattringer:2012ap,Mercado:2013ola,Mercado:2013yta,Kloiber:2014qea,Gattringer:2015nea,Gattringer:2014nxa},
which, however, is irrelevant for a numerical simulation.

In these notes we apply the same dualization techniques as in 
\cite{Gattringer:2012df,Gattringer:2012ap,Mercado:2013ola,Mercado:2013yta,Kloiber:2014qea,Gattringer:2015nea,Gattringer:2014nxa}
to a simple system with a topological term, the U(1) gauge-Higgs model in two 
dimensions. We show that a dual representation with only real and positive contributions is possible for arbitrary
values of the vacuum angle $\theta$ and implement a suitable Monte Carlo algorithm. This constitutes the first example of a complete 
solution of the complex action problem coming from a topological term. First results from the dual simulation of the U(1) gauge-Higgs 
model with non-zero $\theta$ were presented in 
\cite{Kloiber:2014qea}.

The two-dimensional U(1) gauge-Higgs model is not only interesting as a testbed for the dual approach, 
but also provides interesting physics. 
In its Higgs phase directly related to the phenomenological Ginzburg-Landau model of 
superconductivity \cite{Ginzburg:1950sr} it exhibits well-localized multi-vortex solutions
with non-trivial integer topological charge  
\cite{Nielsen:1973cs,Bogomolny:1976tp,Bogomolny:1975de,Jaffe:1980mj}. On the lattice
the model and its topological effects have been studied in two  
\cite{Bunk:1982rk,Grunewald:1986ib} as well as in four
dimensions \cite{Ranft:1982hf,Damgaard:1987ec}. 

Here we are going to demonstrate that the topological features of the model in two dimensions can be further 
explored with the new techniques which now allow for simulations at arbitrary $\theta$.

\section{U(1) gauge Higgs model with a topological term}

\subsection{Continuum formulation}

The Euclidean continuum action for the 2-dimensional U(1) gauge Higgs model reads
\begin{align}
S[A,\phi]  \; &= \; \int \! d^{\,2}x \Big[ (D_\mu(x) \phi(x))^* D_\mu(x) \phi(x) \\
&+m^2|\phi(x)|^2+\lambda |\phi(x)|^4 
+ \frac{\beta}{4} F_{\mu\nu}(x) F_{\mu\nu}(x) \Big]~. \nonumber
\end{align}
Here  $D_\mu(x) = \partial_\mu + i A_\mu(x)$  denotes the usual covariant derivative and 
$F_{\mu\nu}(x) = \partial_\mu A_\nu(x) - \partial_\nu A_\mu(x)$
is the field strength tensor for the U(1) gauge field $A_\mu(x)$. The matter fields are described by a complex scalar field $\phi(x)$.
 
The topological charge for U(1) gauge fields in two dimensions is given by 
\begin{equation}
Q[A] \! =  \! \frac{1}{4\pi} \int \! d^2x \, \epsilon_{\mu\nu} F_{\mu\nu}(x) \! = \! \frac{1}{2\pi} \int \! d^2x ~F_{12}(x) \; .
\label{topocont}
\end{equation}
The topological term is added with a vacuum angle $\theta$ to the action and the partition function reads
\begin{equation}
Z \; = \; \int \! \mathcal{D}[A]\mathcal{D}[\phi]  \; e^{-\, S[A,\phi] \,  - \, i \theta Q[A]} \; .
\end{equation}

\subsection{Lattice formulation}

On the lattice the partition function is given by
\begin{equation}
Z  \; =  \; \int \! \mathcal{D}[U]~\mathcal{D}[\phi] \; e^{-\, S_G[U] \, - \, S_M[U,\phi] \, -i\,\theta \, Q[U]} \; .
\label{zconventional}
\end{equation}
For the gauge part we use the standard Wilson gauge action
\begin{equation}
S_G[U] \; = \; - \, \frac{\beta}{2} \sum_x \big[\, U_{x,p} \, + \, U_{x,p}^{\;*}\, \big] ~,
\end{equation}
where the sum runs over the sites $x$ of a 2-dimensional \(N_t \times N_s\) 
lattice with periodic boundary conditions. The plaquette variable $U_{x,p}$ is built from the 
U(1)-valued link variables $U_{x,\mu} \, , \; \mu = 1,2$, and is given by
$U_{x,p} \; = \; U_{x,1} \, U_{x+\hat{1},2} \, U_{x+\hat{2},1}^{\,*} \, U_{x,2}^{\,*} \; .$ 
The matter part of the action reads
\begin{align}
S_M[U,\phi] \; = \; & \sum_x \Big[ \kappa |\phi_x|^2+\lambda|\phi_x|^4 \\
& -\sum_\mu(\phi^*_x U_{x,\mu} \phi_{x+\hat{\mu}} + \phi_x U_{x,\mu}^* \phi^*_{x+\hat{\mu}} )\Big]~, \nonumber
\end{align}
with a mass parameter $\kappa=4+m^2$.

We stress at this point that in the literature (see, e.g., \cite{jansen:85}) a different nomenclature can be found, where 
the couplings have a slightly different meaning (we add primes to distinguish them from our 
couplings):  $\kappa^\prime$ is the factor in front of the nearest neighbor terms, $\lambda^\prime$ is
the factor of a shifted quartic term of the form $(|\phi_x|^2-1)^2$, and the quadratic term comes with a
factor of 1. The connection between the two conventions is given by the transformations
\begin{equation}
\lambda = \frac{\lambda^\prime}{\kappa^{\prime \; 2}} \; , \; \;
\kappa = \frac{1 - 2 \lambda^\prime}{\kappa^\prime} \; , \; \;
\phi_x = \frac{1}{\sqrt{\kappa^\prime}} \phi_x^\prime \; , 
\end{equation}
and by dropping an irrelevant term. In the convention \cite{jansen:85} the limit $\lambda^\prime \rightarrow \infty$ freezes the radial
mode to $|\phi_x^\prime| = 1$ and one expects a Kosterlitz-Thouless transition when varying $\kappa^\prime$. In our convention
this corresponds to following the line $\lambda = (1- \kappa^\prime \kappa)/2\kappa^{\prime \; 2}$ for $\lambda \rightarrow \infty$ 
(compare also the $\kappa$-$\lambda$ phase diagram in Fig.~\ref{fig_kl_phase_diagram} below).

On the lattice the topological charge can be discretized with, e.g., the ``field-theoretical definition'',
\begin{equation}
Q[U] \; = \; \frac{1}{i 4 \pi} \sum_x \big[ \; U_{x,p} \,  - \, U_{x,p}^{\; *} \, \big] \; ,
\label{topcharge}
\end{equation} 
which reproduces (\ref{topocont}) in the continuum limit. For the naive continuum limit this can be seen, by setting 
$U_{x,\mu} = e^{i A_{x,\mu}}$ and noting that
\begin{equation}
U_{x,p} = e^{i ( A_{x,1} + A_{x + \hat{1},2} - A_{x+\hat{2},1} -A_{x,2} )}  = 1  + i F_{12}(x) + \ldots
\end{equation}
The measures  $\mathcal{D}[U]$ and $\mathcal{D}[\phi]$ in the lattice path integral are defined in the usual compact way, i.e., as 
product measures over all degrees of freedom on the lattice, 
\begin{equation}
\int\!\mathcal{D}[U] \; = \;  \prod_{x,\mu} \int_{-\pi}^{\pi} \frac{d A_{x,\mu}}{2\pi}~ , ~~  
\int\!\mathcal{D}[\phi] \; = \; \prod_{x} \int_{\mathds{C}} \frac{d \phi_x}{2\pi} \; .
\end{equation}
Putting things together we can write the partition sum as
\begin{align}
Z  \; &= \; \int\! \mathcal{D}[U] \; e^{ \, \eta \sum_x U_{x,p} \; + \; \overline{\eta} \sum_x U_{x,p}^{\, *}} \; Z_M[U] ~, \nonumber \\
Z_M[U]  \; &= \; 
\int\! \mathcal{D}[\phi] \; e^{- \, S_M[U,\phi]} \; ,
\label{latticeZ}
\end{align}
where $Z_M[U]$ is the partition sum of the matter fields in the gauge background. For a convenient notation of the terms that combine 
the gauge action and the topological charge we defined 
\begin{equation}
\eta\equiv\frac{\beta}{2}-\frac{\theta}{4\pi}~, ~~~~~ \bar{\eta}\equiv\frac{\beta}{2}+\frac{\theta}{4\pi}~.
\label{etadef}
\end{equation}

It is obvious, that the conventional representation 
(\ref{latticeZ}) of the lattice model is not suitable for a Monte Carlo simulation at $\theta \neq 0$, since then $\eta \neq \overline{\eta}$
and the Boltzmann factor is complex. In the next subsection we show that this problem is overcome by mapping the partition sum 
to dual variables.
 
\subsection{Dual representation}

By expanding the Boltzmann factors containing the nearest neighbor terms in $Z_M[U]$ one can exactly map the partition 
sum of the matter fields into a  dual form  \cite{Gattringer:2012df,Gattringer:2012ap,Mercado:2013ola,Mercado:2013yta},
where the new degrees of freedom for the matter fields are loops  dressed with the link variables $U_{x,\mu}$.
For the gauge fields, one proceeds in a similar way \cite{Mercado:2013ola,Mercado:2013yta}, expanding the Boltzmann factor, rearranging terms and integrating 
out the link variables. A more detailed account of this exact transformation of the partition function $Z$ 
into its dual form is provided in the appendix. 

In terms of the dual variables the partition function (\ref{latticeZ}) is given by
\begin{widetext}
\begin{eqnarray}
Z  & = & \sum_{\{l,\bar{l},p,\bar{p}\}} \Bigg[ \prod_{x,\mu} ~\frac{1}{(|l_{x,\mu}|+\bar{l}_{x,\mu})!~ \bar{l}_{x,\mu}!} \Bigg] \Bigg[ \prod_x P(n_x) \Bigg]
\Bigg[ \prod_x ~\frac{\eta^{(|p_x|+p_x)/2+\bar{p}_x} ~ \overline{\eta}^{\,(|p_x|-p_x)/2+\bar{p}_x}}{(|p_x|+\bar{p}_x)!~\bar{p}_x!} \Bigg]
\nonumber \\
& \times &
\Bigg[ \prod_x ~ \delta \Big(\sum_\mu \big[l_{x,\mu}-l_{x-\hat{\mu},\mu}\big]\Big) 
~ \delta(p_x-p_{x-\hat{2}}+l_{x,1})~\delta(p_{x-\hat{1}}-p_x+l_{x,2})\Bigg]~.
\label{z_dual}
\end{eqnarray}
\end{widetext}
In the dual representation the partition function is a sum over the set $\{l,\bar{l},p,\bar{p}\}$  of all configurations of the integer valued 
dual variables 
\begin{equation}
l_{x,\mu}\, , \; p_x \in \mathds{Z}~, ~~~~~ \bar{l}_{x,\mu} \, , \; \bar{p}_x \in \mathds{N}_0 \; ,
\end{equation}
which are assigned to the links ($l_{x,\mu}$ and $\bar{l}_{x,\mu}$) or the plaquettes  ($p_x$ and $\bar{p}_x$) of the lattice.
For each configuration there is a real and positive weight factor which consists of the terms in the first line of 
(\ref{z_dual}), where we have introduced the following abbreviations
\begin{align}
\label{p_n_def}
P(n_x) &\equiv \int_0^\infty \!\! dr ~ r^{n_x+1} ~ e^{-\kappa r^2 - \lambda r^4}~, \\
n_x &\equiv \sum_\mu \big[ \, |l_{x,\mu}|+|l_{x-\hat{\mu},\mu}|+2(\, \bar{l}_{x,\mu}+\bar{l}_{x-\hat{\mu},\mu}\,) \,\big] \; . \nonumber
\end{align}
A subset of the dual variables, i.e., the $l_{x,\mu}$ and the $p_x$, are subject to constraints, which are collected in the second line
of (\ref{z_dual}). Here the $\delta$ denote Kronecker deltas, i.e., $\delta(n) \equiv \delta_{n,0}$. The constraints for the $l_{x,\mu}$
are a discretized version of $\nabla \vec{l}_x = 0$ and thus imply the conservation of $l$-flux at every site $x$.
The remaining constraints are associated with the fluxes along the links of the lattice: Here the flux at a link introduced by a non-trivial
$p_x$ has to be compensated by an oppositely oriented flux from a neighboring plaquette or by $l$-flux from link variables $l_{x,\mu}$.
The combinations of all constraints gives rise to admissible configurations that contribute to the partition sum, which consist of closed loops of
$l$-flux which are filled with occupied plaquettes such that at each link the total flux is zero. These admissible configurations are a 
natural reduction of the admissible configurations that are discussed in more detail for the 4-dimensional case in \cite{Mercado:2013ola,Mercado:2013yta}.

We close this subsection with remarking that the weight factors in the dual representation (\ref{z_dual}) are always real, but become 
negative when either $\eta < 0$ or $\overline{\eta} < 0$. In practice this is, however, an irrelevant region of the parameters, since we
are interested in the limit $\beta \rightarrow \infty$ where one approaches the continuum limit (see below). Thus we can restrict ourselves to 
the parameter region \(\beta>\theta/2\pi\) where both $\eta$ and $\overline{\eta}$ are positive. 

\vfill

\subsection{Observables and their dual representation}
\label{sec_observables}

For the analysis here, we focus on studying the behavior of various bulk observables (for a calculation of propagators in the dual 
picture see, e.g., \cite{Gattringer:2012df,Gattringer:2012ap}). In particular we consider:

Square of the absolute field and its susceptibility:
\begin{equation}
\langle |\phi|^2 \rangle \; \equiv \; \frac{-1}{N_s\,N_t} \frac{\partial}{\partial \kappa} \ln Z~,~
\chi_{\phi} \; \equiv \; \frac{1}{N_s\,N_t} \frac{\partial^2}{\partial \kappa^2} \ln Z \; .
\end{equation}

Plaquette and plaquette susceptibility:
\begin{equation}
\langle \mbox{Re}\, U_p \rangle \; \equiv \; \frac{1}{N_s \, N_t} \frac{\partial}{\partial \beta} 
\ln Z~,~\chi_{p} \; \equiv \; \frac{1}{N_s \, N_t} \frac{\partial^2}{\partial \beta^2} \ln Z \; .
\end{equation}

Topological charge density and topological charge susceptibility:
\begin{equation}
\langle q \rangle \; \equiv \;  \frac{-1}{N_s \, N_t} \frac{\partial}{\partial \theta} \ln Z~,~\chi_{t} 
\; \equiv \; \frac{-1}{N_s \, N_t} \frac{\partial^2}{\partial \theta^2} \ln Z \; .
\label{topobs}
\end{equation}

The dual expressions of these observables 
can be obtained by evaluating the derivatives of $\ln Z$ using the dual representation of the partition function $Z$. We give 
two examples for dual expressions of observables, the somewhat simpler field expectation value,
\begin{align}
\langle |\phi|^2 \rangle \! &= \! \frac{-1}{N_s\,N_t} \frac{\partial}{\partial \kappa} \ln Z \! = \! \frac{-1}{N_s\,N_t} \left\langle 
\sum_x \frac{\partial P(n_x)}{\partial \kappa} \frac{1}{P(n_x)} \right\rangle \; \nonumber \\
& = \;  \frac{1}{N_s\,N_t} \left\langle 
\sum_x \frac{P(n_x +2 )}{P(n_x)} \right\rangle \; ,
\label{dualobs1}
\end{align}
and the slightly more involved expression for the topological charge susceptibility,
\begin{widetext}
\begin{eqnarray}
\chi_{t} &\! = \! &  \frac{-1}{ N_s N_t} \Bigg[ \frac{1}{(4\pi\eta)^2} \left\langle \left[ \frac{|\mathcal{S}|}{2}\! +\! \frac{\mathcal{S}}{2} \!+ \!
\overline{\mathcal{S}} \right]\!\left[ \frac{|\mathcal{S}|}{2} \!+\! \frac{\mathcal{S}}{2} \!+\! \overline{\mathcal{S}} \!-\!1\right]\right\rangle 
\!-\!  \frac{1}{8\pi^2\eta \bar{\eta}} \left\langle \left[ \frac{|\mathcal{S}|}{2} \!+\! \frac{\mathcal{S}}{2} \!+\! \overline{\mathcal{S}} \right] \!
\left[ \frac{|\mathcal{S}|}{2} \!-\! \frac{\mathcal{S}}{2} \!+\! \overline{\mathcal{S}}\right]\right\rangle
\label{dualobs2}
\\
& \!+\! &\frac{1}{(4\pi\bar{\eta})^2} \left\langle \left[ \frac{|\mathcal{S}|}{2}\! -\! \frac{\mathcal{S}}{2} \!+ \!
\overline{\mathcal{S}} \right]\left[ \frac{|\mathcal{S}|}{2} \!- \!\frac{\mathcal{S}}{2} \!+\! \overline{\mathcal{S}} \!-\!1\right]\right\rangle
\!- \! \left\langle \frac{1}{4\pi\bar{\eta}}\left[ \frac{|\mathcal{S}|}{2} \!- \!\frac{\mathcal{S}}{2} \!+ \!\overline{\mathcal{S}} \right] -
\frac{1}{4\pi\eta}\left[ \frac{|\mathcal{S}|}{2} \! +\! \frac{\mathcal{S}}{2} \!+\! \overline{\mathcal{S}}\right] \right\rangle^2 \Bigg] ,
\nonumber 
\end{eqnarray}
\end{widetext}
where we use the abbreviations
$ |\mathcal{S}| \equiv \sum_x |p_x|, \; \mathcal{S} \equiv \sum_x p_x$ and $\overline{\mathcal{S}} \equiv \sum_x \bar{p}_x$ for various 
sums of the plaquette occupation numbers $p_x$ and $\bar{p}_x$. The expectation values on the right-hand sides of (\ref{dualobs1}) and
(\ref{dualobs2}) are understood as expectation values in the dual representation. In a similar way all bulk observables defined 
above can be obtained as weighted moments of the dual variables. 

\section{Tests in pure gauge theory and dual Monte Carlo updates}

\subsection{Pure U(1) gauge theory: Semi-analytical results and continuum limit}
\label{sec_puregauge}

If we neglect the matter fields (''quenched case''), the model can be solved (semi-) analytically, i.e., we obtain for the partition 
function a simple and fast converging sum, which can be evaluated efficiently to arbitrary precision.

For the pure gauge system with topological term the partition function reduces to
\begin{align}
Z  \; = \; &\sum_{\{p\}} \Bigg[ \prod_x \sum_{\bar{p}_x=0}^\infty \frac{\big(\sqrt{\eta \overline{\eta}}\big)^{\bar{p}_x+|p_x|}}{(|p_x|+\bar{p}_x)! ~ 
\bar{p}_x!} \Bigg] \Bigg[ \prod_x \left(\sqrt{\frac{\eta}{\overline{\eta}}}\,\right)^{p_x} \Bigg] \nonumber \\
& \times ~
\Bigg[ \prod_x \delta(p_x-p_{x-\hat{2}}) ~ \delta(p_{x-\hat{1}} - p_x) \Bigg],
\end{align}
where the sums over the \(\bar{p}_x\) in the first parentheses are well known and yield the modified Bessel functions ${I}_n(x)$.
We thus obtain for the partition sum the expression
\begin{equation}
\begin{split}
Z \; = \; &\sum_{\{p\}} \Bigg[\prod_x  {I}_{|p_x|}\big(2\sqrt{\eta \overline{\eta}} \,\big) \left(\sqrt{\frac{\eta}{\overline{\eta}}}\,\right)^{p_x} \Bigg] \\
& \times ~ \Bigg[ \prod_x \delta(p_x-p_{x-\hat{2}}) ~ \delta(p_{x-\hat{1}} - p_x) \Bigg].
\label{zquench1}
\end{split}
\end{equation}
In the quenched case we have no matter flux for saturating the constraints at the links. Thus the Kronecker deltas in (\ref{zquench1})
force the plaquette occupation numbers $p_x\in\mathds{Z}$ to have the same value $q$ at each lattice site such that all fluxes
along the links cancel.  Hence every configuration that obeys all constraints can be labeled by a single integer $q$ 
and $p_x = q \; \forall x$. This simplifies the partition sum to 
\begin{equation}
Z \; = \; \sum_{q=-\infty}^{+\infty} \left[ \; {I}_{|q|}\big(2\sqrt{\eta \overline{\eta}}\big) 
\left(\sqrt{\frac{\eta}{\overline{\eta}}}\,\right)^{\!\!q} \; \right]^{ N_s  N_t}.
\label{Zgauge}
\end{equation}
The modified Bessel functions $I_n$ decay faster than exponentially with the index $n$ and thus the series (\ref{Zgauge}) converges 
rapidly. It is straightforward to evaluate it numerically with Mathematica and the results we show below were obtained in this way.

Eq.~(\ref{Zgauge}) nicely illustrates how the vacuum angle $\theta$ influences the physics in our system. For $\theta < 0$ we 
have $\eta > \overline{\eta}$ and thus the term $\big(\sqrt{\eta/\overline{\eta}}\big)^{q}$ enhances configurations
with $q > 0$. Configurations with $q_x = q \; \forall x$ correspond to configurations of constant electric flux and via the term 
$\big(\sqrt{\eta/\overline{\eta}}\big)^{q}$ the vacuum angle allows to introduce such flux in the system. 

We can also use (\ref{Zgauge}) for a first assessment of the continuum limit. In particular it is interesting to study how well the field theoretical 
lattice definition of the topological charge (\ref{topcharge}) can reproduce the expected $2\pi$-periodicity of observables as a function
of the vacuum angle $\theta$. This is not a priori clear, since the definition (\ref{topcharge}) does not guarantee an integer valued topological 
charge -- this is expected only in the continuum limit.
The continuum limit is approached via
\begin{equation}
\beta \rightarrow \infty ~~ N_s, ~~ N_t \rightarrow \infty ~~\mbox{with} ~~ \frac{\beta}{N_s ~ N_t} \; = \; \mbox{const.} 
\label{contlimit}
\end{equation}
Dimensional analysis yields $[\beta]=L^2$ which implies that the continuum limit (\ref{contlimit}) 
corresponds to keeping a fixed physical volume. One expects that in the fixed-volume continuum 
limit observables become $2 \pi$-periodic in $\theta$. 

\begin{figure*}[t]
\begin{center}
\includegraphics[height=6.3cm,type=pdf,ext=.pdf,read=.pdf]{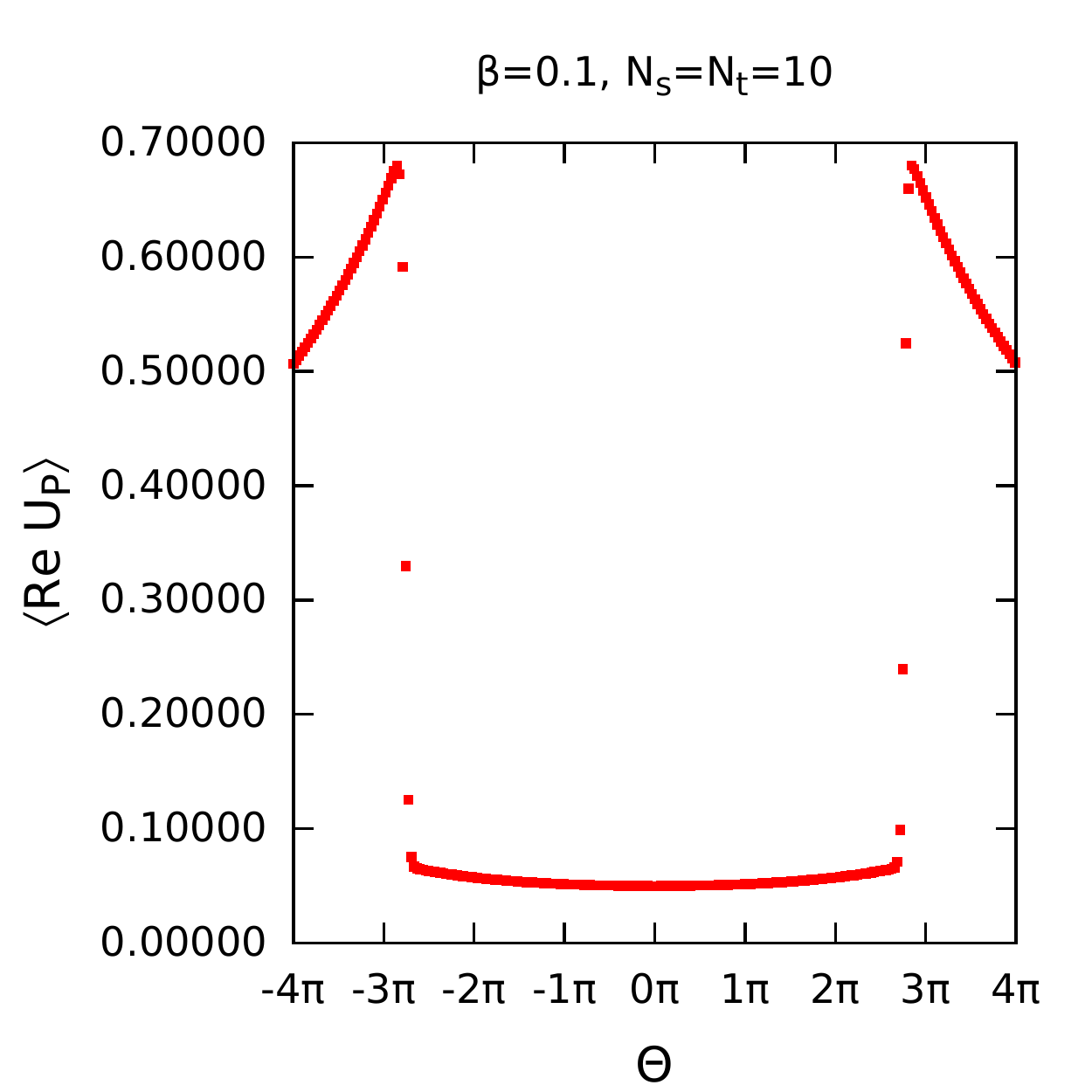}
\includegraphics[height=6.3cm,type=pdf,ext=.pdf,read=.pdf]{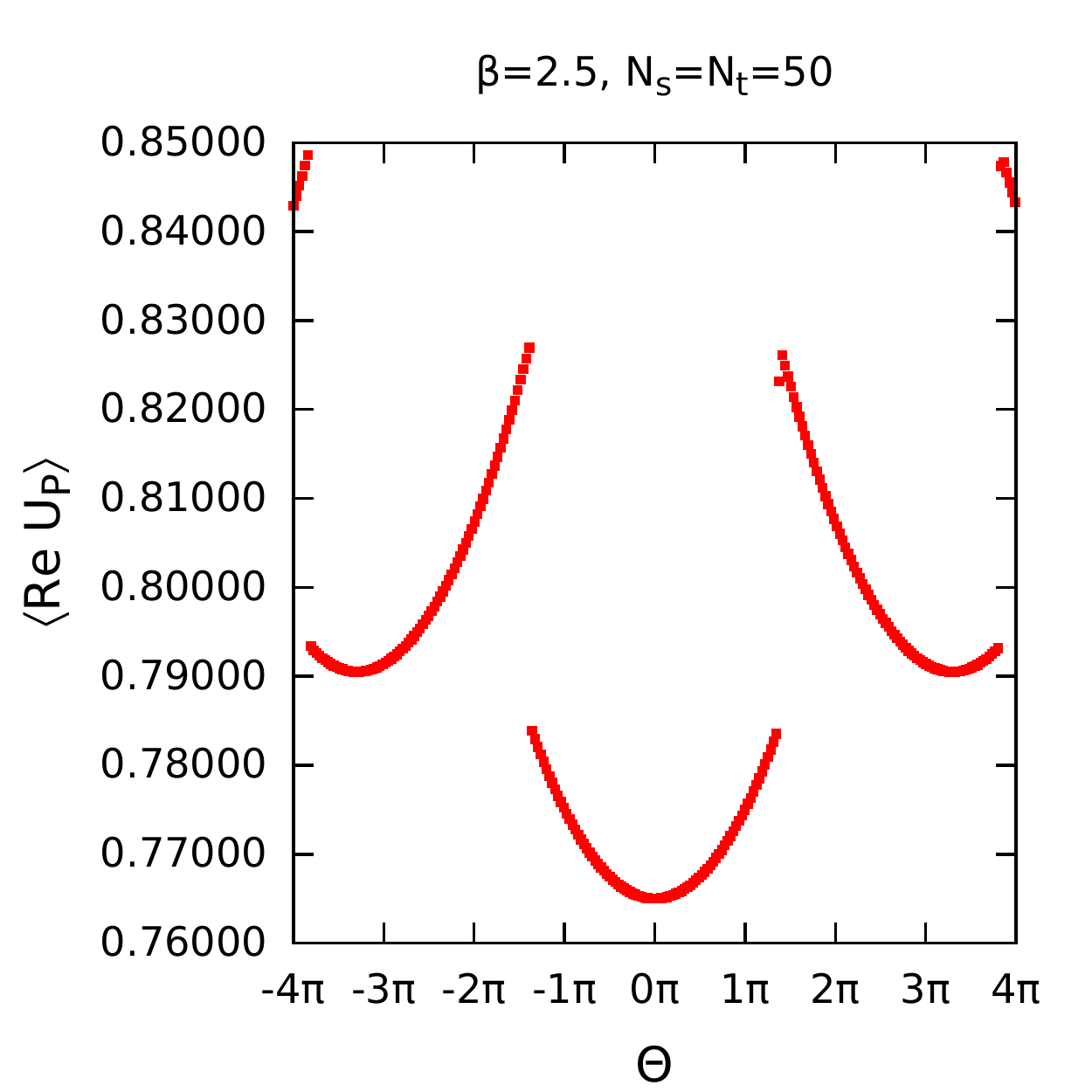}

\includegraphics[height=6.3cm,type=pdf,ext=.pdf,read=.pdf]{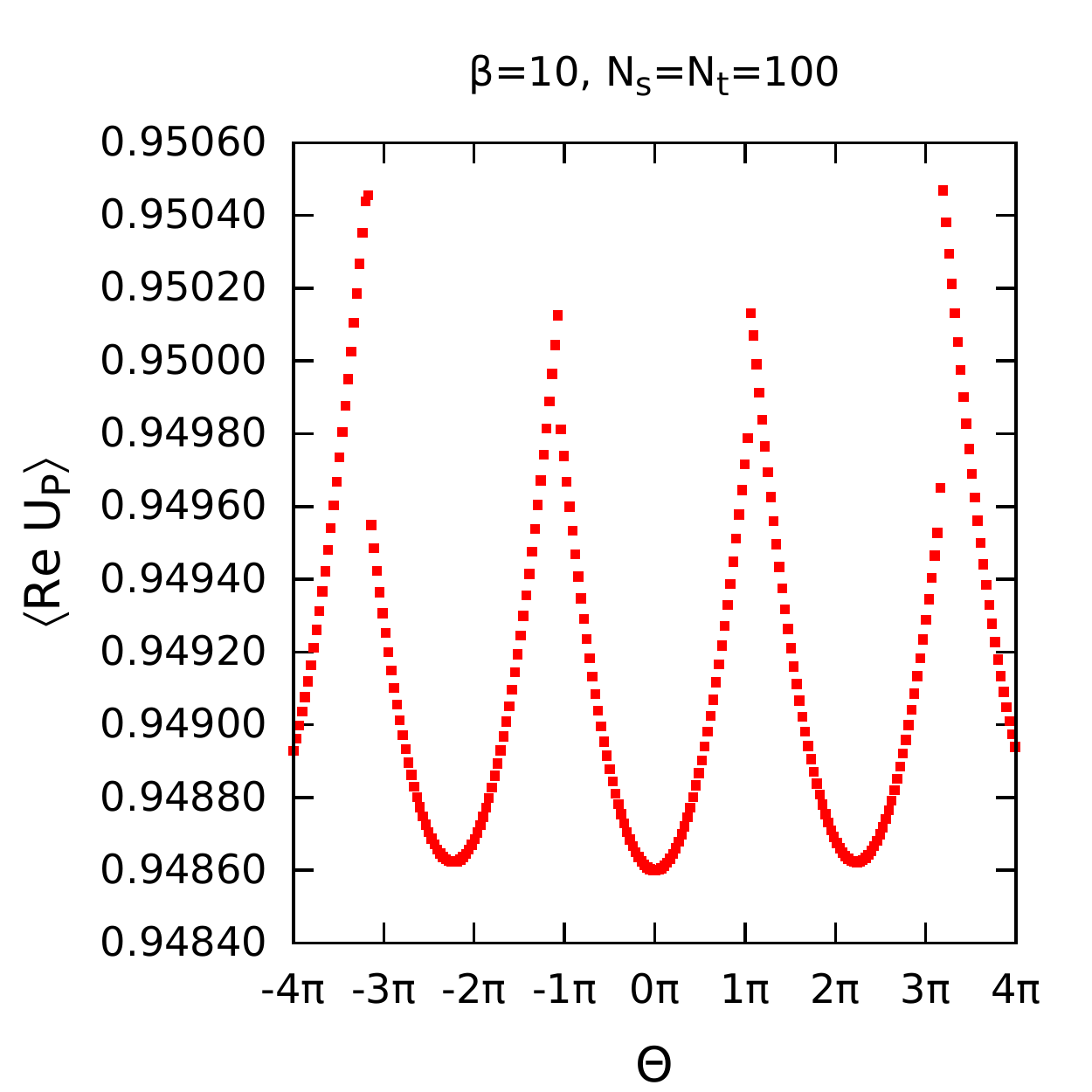}
\includegraphics[height=6.3cm,type=pdf,ext=.pdf,read=.pdf]{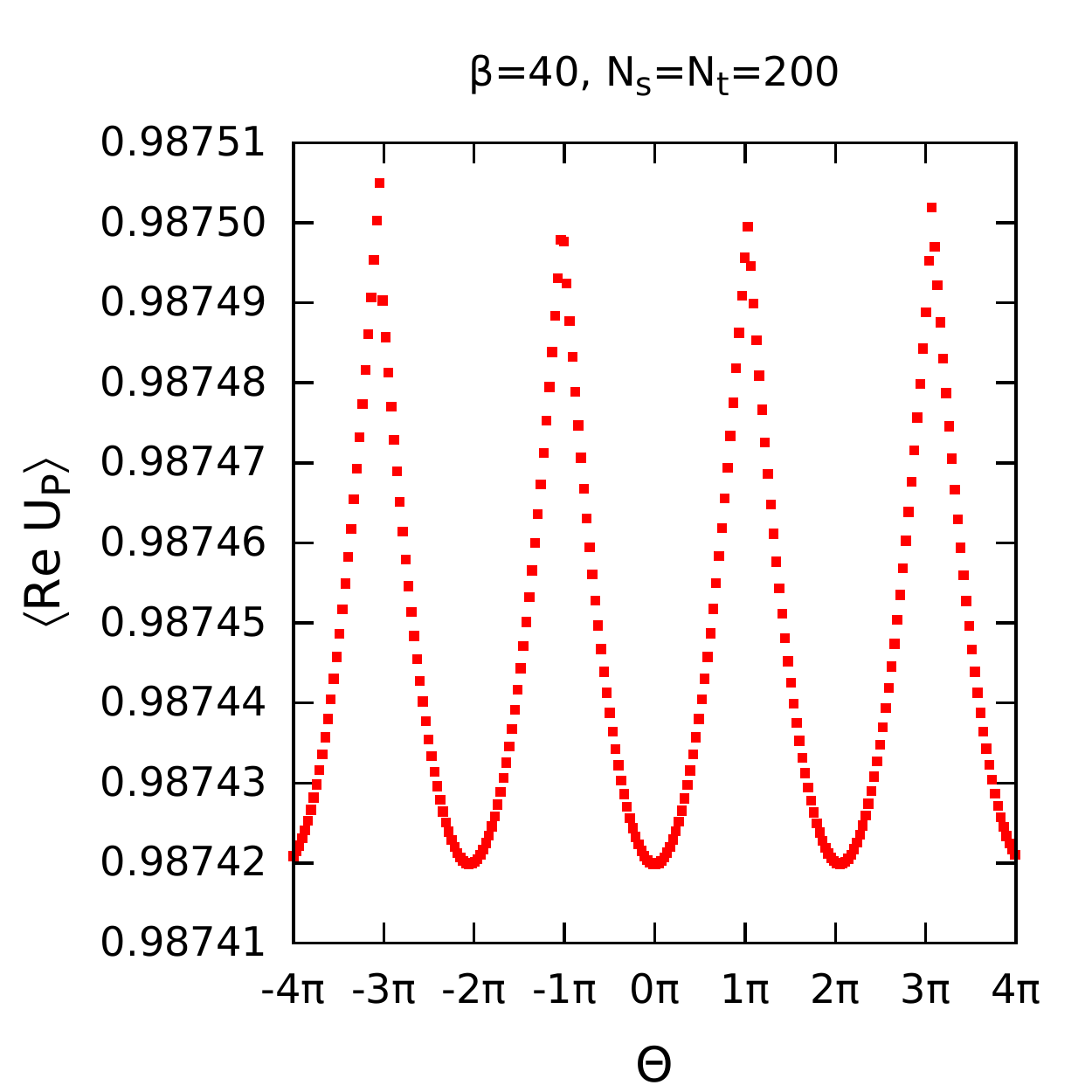}
\end{center}
\caption{Plaquette expectation value \(\langle\mbox{Re} \, U_p \rangle\) of the pure gauge theory versus the vacuum angle \(\theta\). We 
show the approach to the continuum limit using \(\beta = 0.1, 2.5, 10.0\) and $40.0$ at fixed \(\beta/N_s N_t = 0.001\). The plots 
nicely demonstrate how the observable becomes \(2\pi\)-periodic in \(\theta\) when approaching the continuum limit.}
\label{fig1}
\end{figure*}

We studied the $\theta$-dependence of various observables and as an example in Fig.~\ref{fig1} we show 
the plaquette expectation value as a function of $\theta$ on a sequence of lattices that approach the 
continuum limit (\(\beta = 0.1, 2.5, 10.0\) and \(40.0\) at fixed \(\beta/N_s N_t = 0.001\)). The results were
obtained by evaluating $\langle \, \mbox{Re} \, U_p \rangle \, = \, \frac{\partial}{\partial \beta} \ln Z / N_t N_s$ for the 
pure gauge partition sum as given in (\ref{Zgauge}). The tests documented in Fig.~\ref{fig1} show that
indeed \(2\pi\)-periodicity is recovered and the field theoretical definition (\ref{topcharge}) is well suited to study 
the continuum limit of the model with the topological term. 

\subsection{Dual Monte Carlo simulation}

Before we come to the presentation of the results for the full model, let us discuss the Monte Carlo update strategy used for the dual 
representation. In the dual representation (\ref{z_dual}) the dynamical degrees of freedom are the integer valued plaquette and matter flux 
variables, $p_x, \bar p_x$ and $l_{x,\mu}, \bar l_{x,\mu}$. The dual variables  $\bar p_x$ and $\bar l_{x,\mu}$ are not subject to any 
constraints and they can be updated independently in the usual way with a local Monte Carlo update scheme. More demanding are the
variables $p_x$ and $l_{x,\mu}$ which have to obey the constraints of conserved $l$-flux at each site and a vanishing of the 
combined $p$- and $l$-flux at each link of the lattice. 

Although it is possible to find a generalization of the worm strategy \cite{Prokof'ev:2001zz} to abelian gauge-Higgs models in arbitrary dimensions
\cite{Mercado:2013ola,Mercado:2013yta}, for the 2-dimensional model studied here we use a simpler local update for the dual variables. 
The update strategy consists of two types of updates:  

\begin{enumerate}
\item A \textit{local plaquette/link update}: For a lattice site $x$ we randomly choose $\Delta_x = \pm 1$ and propose to change
\begin{equation}
p_x \rightarrow p'_x = p_x + \Delta_x~,
\label{mc1}
\end{equation}
\begin{equation}
\begin{split}
&l_{x,1} \rightarrow l'_{x,1}-\Delta_x~,~~~l_{x+\hat{1},2} \rightarrow l'_{x+\hat{1},2}-\Delta_x~, \\
&l_{x+\hat{2},1} \rightarrow l'_{x+\hat{2},1}+\Delta_x~,~~~l_{x,2} \rightarrow l'_{x,2}+\Delta_x \; .
\end{split}
\label{mc2}
\end{equation}
The change is accepted with a Metropolis step.

\item A \textit{global winding gauge update}: A $\Delta_x = \pm 1$ is chosen randomly and we propose to change
\begin{equation}
p_x \rightarrow p'_x = p_x + \Delta ~~~~~ \forall x \; .
\label{mc3}
\end{equation}
Again the change is accepted with a Metropolis step.
\end{enumerate}
It is easy to see that these updates leave the constraints intact and that 
this procedure is ergodic. To be precise, the plaquette update alone is already ergodic, but mixing sweeps of the 
local plaquette/link update with global winding gauge updates considerably reduces the auto-correlation of observables
related to the topological charge. 
For the results we show here, we typically use $5 \times 10^4$ such combined  sweeps for equilibration followed by $10^6$ measurements 
separated by
5 combined sweeps for decorrelation. The error bars we show are statistical errors determined with a jackknife analysis.
 
The performance of the Monte Carlo updates (\ref{mc1}), (\ref{mc2}) was studied
for the 4-d case in great detail in \cite{Mercado:2013yta}, and qualitatively the performance behavior is the 
same in the 2-d case, which is, however, numerically considerably less demanding.
As remarked, the update (\ref{mc3}) is not necessary for ergodicity,
but helps for decorrelating topological quantities since it corresponds to changing the
topological sector. Thus the acceptance of this update depends exponentially on the volume, 
which is, however, not surprising, since the action for an additional charge
is extensive for the 2-d U(1) case and the exponential volume-dependence is physical (this is different for SU(3) in 4-d).

\section{Numerical results for the full model}

\subsection{Continuum limit and periodicity in the \(\theta\) angle}
\label{subsec_cont_limit}

\begin{figure}[t]
\begin{center}
\includegraphics[width=8cm,type=pdf,ext=.pdf,read=.pdf]{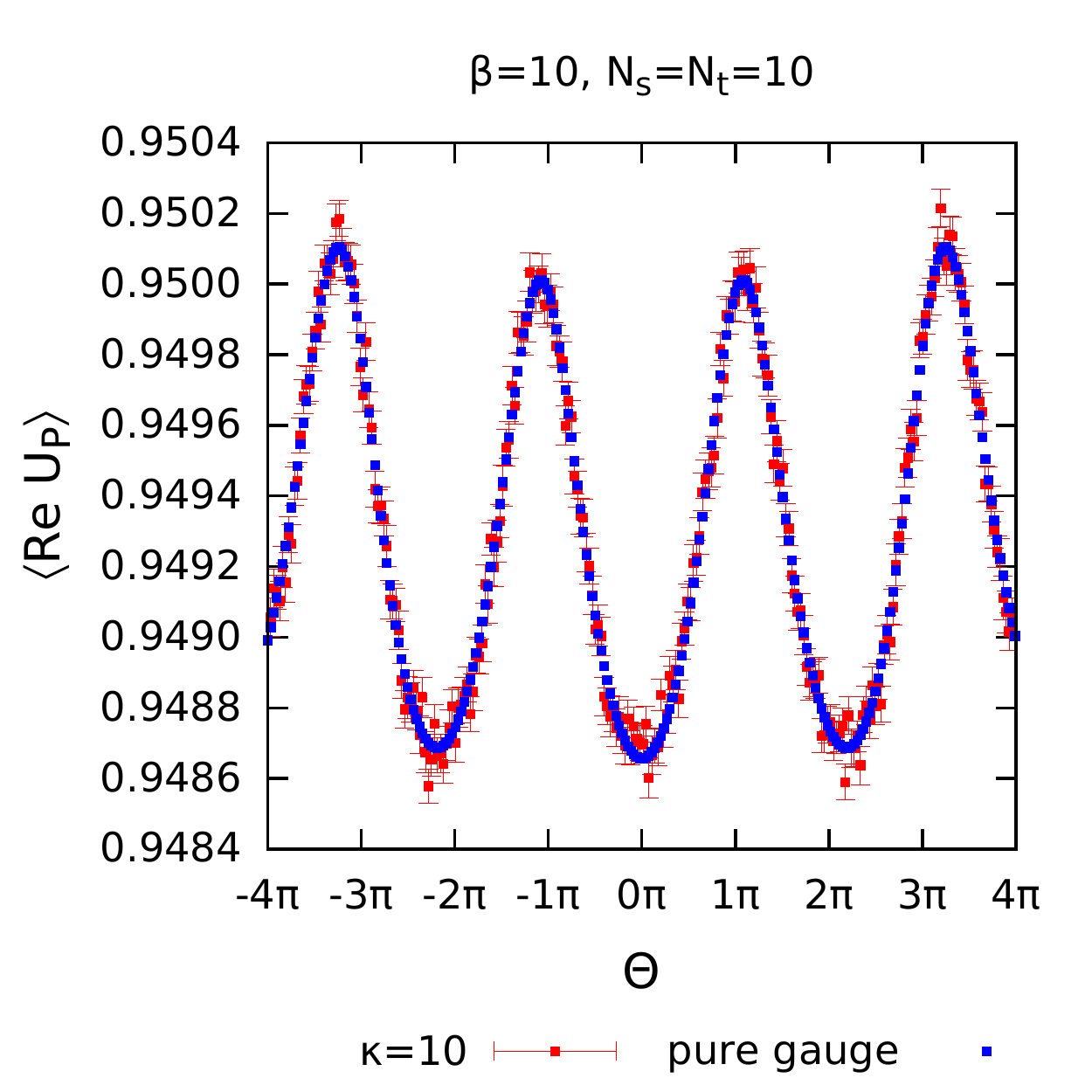}
\end{center}
\caption{Plaquette expectation value \(\langle\mbox{Re} \, U_p \rangle\) of the full model versus the vacuum angle \(\theta\). We show 
data for a large value of \(\kappa = 10.0\) at \(\lambda=1\) \(N_s=N_t=10\). 
We compare the numerical data from the Monte Carlo simulation to the results from pure gauge theory discussed in the previous section.}
\label{fig2}
\end{figure}

Now we study the full model with matter fields. As a first consistency check we discuss a simulation at large $\kappa$, i.e.,  
large mass, where the matter 
fields become static and the results are expected to approach the quenched results from the last section (the quartic coupling 
$\lambda$ is always set to $\lambda = 1$ in this subsection).  Figure \ref{fig2} shows the plaquette 
expectation value as a function of $\theta$ for $\kappa = 10.0$, \(\beta=10.0\) and \(N_s=N_t=10\). It is obvious that, as expected, 
the numerical data from the dual simulation are indeed in very good agreement with the curve obtained from explicitly summing the 
quenched partition function (\ref{Zgauge}).

\begin{figure*}[t!]
\begin{center}
\includegraphics[height=7cm,type=pdf,ext=.pdf,read=.pdf]{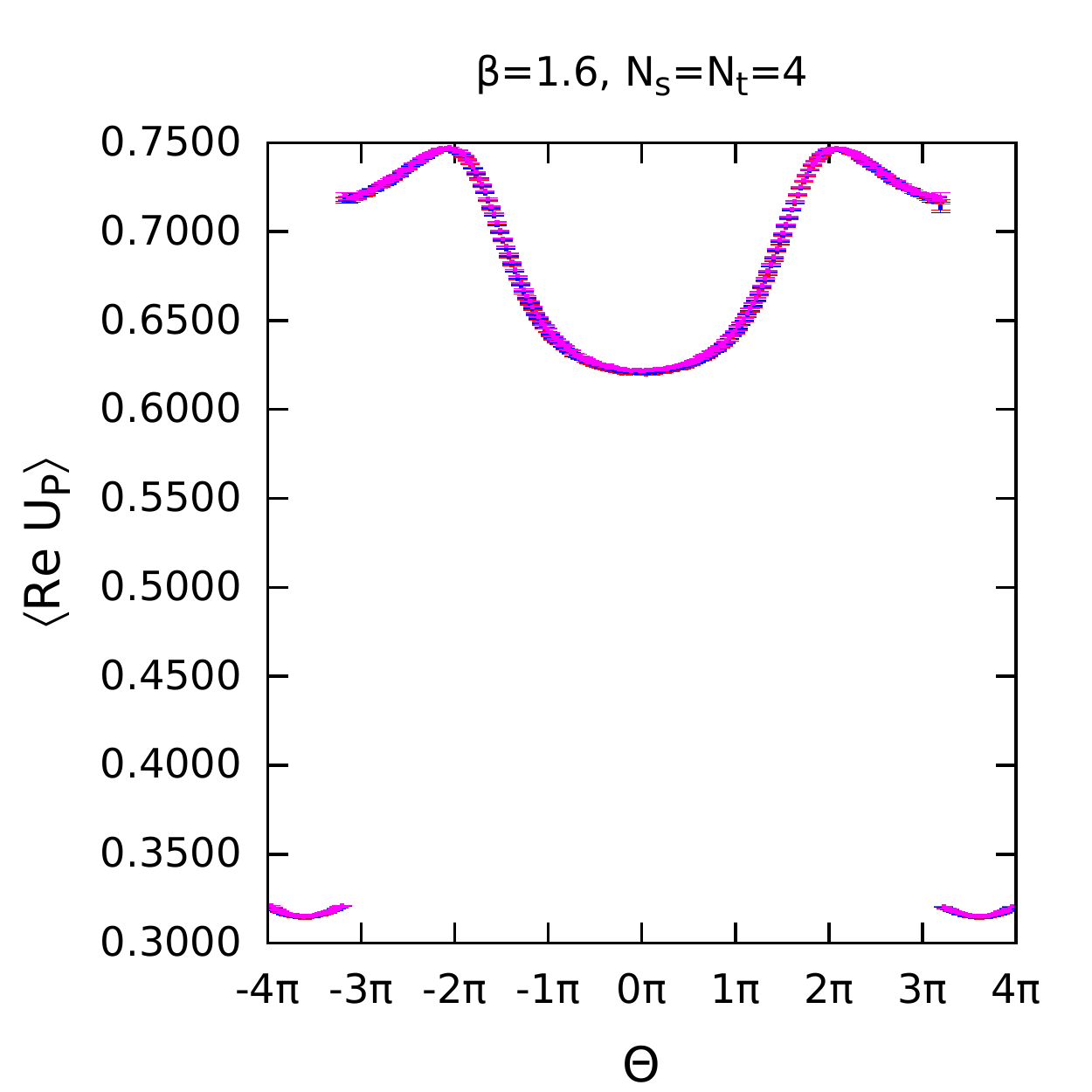}
\hskip5mm
\includegraphics[height=7cm,type=pdf,ext=.pdf,read=.pdf]{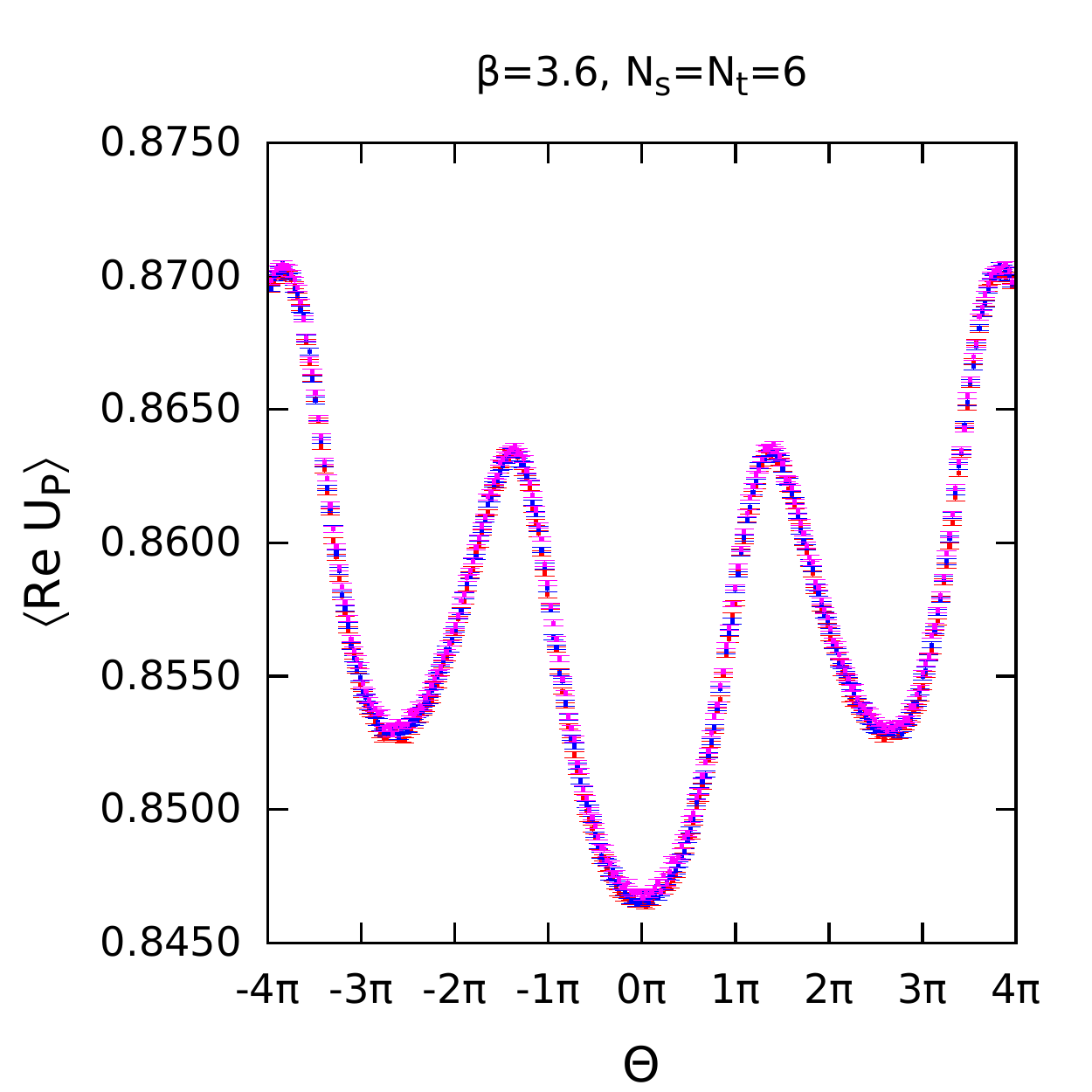}

\includegraphics[height=7cm,type=pdf,ext=.pdf,read=.pdf]{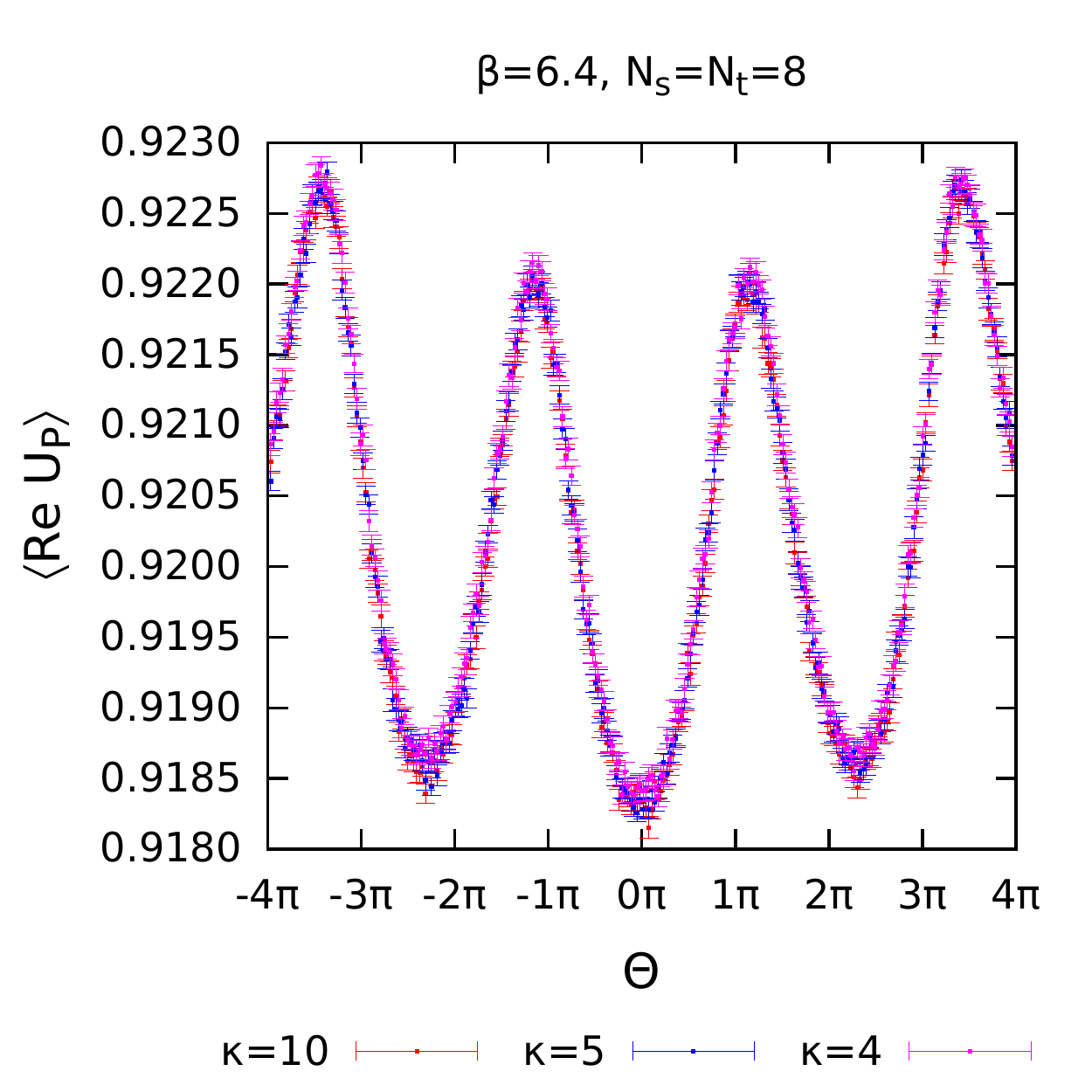}
\hskip5mm
\includegraphics[height=7cm,type=pdf,ext=.pdf,read=.pdf]{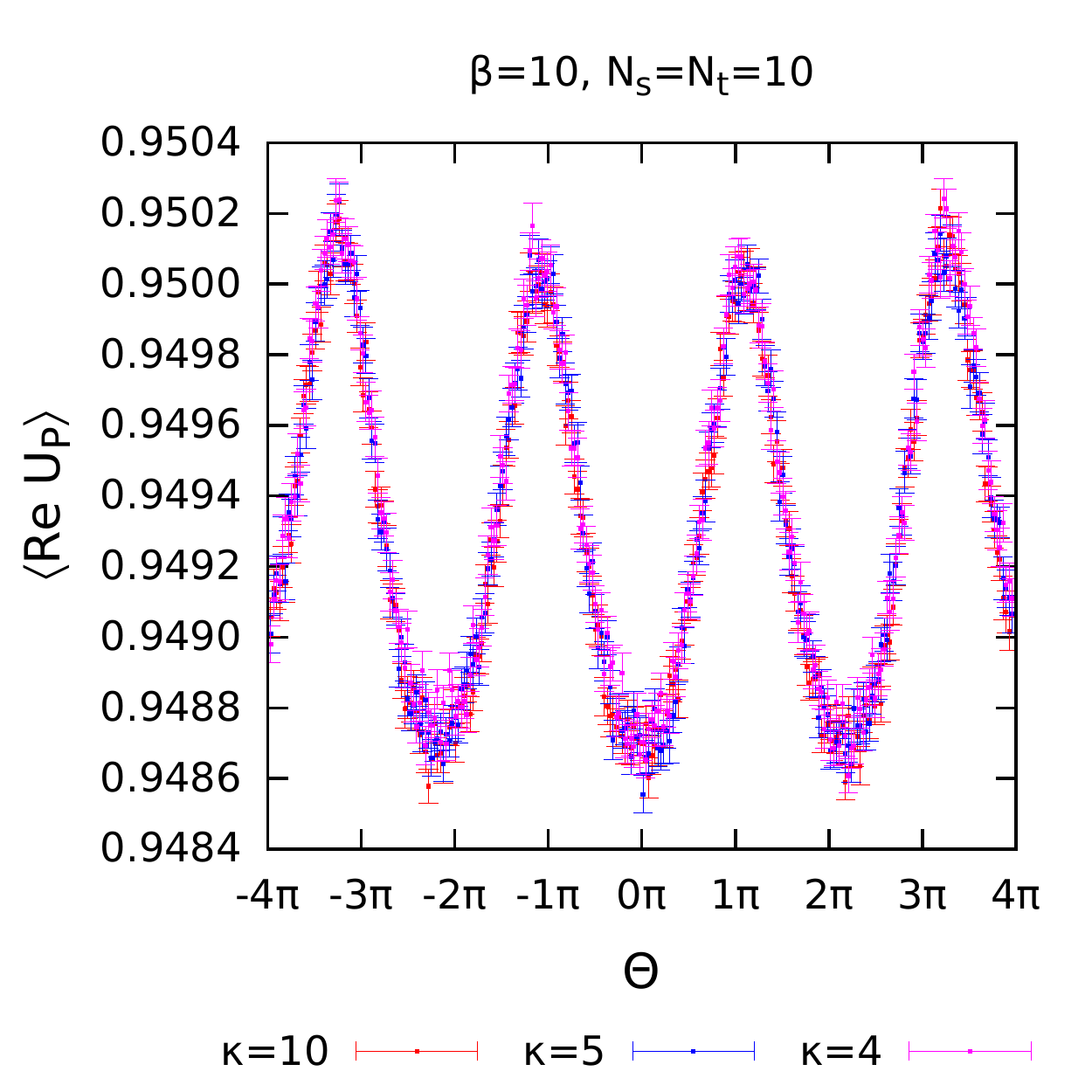}
\end{center}
\caption{Plaquette expectation value \(\langle\mbox{Re} \, U_p \rangle\) of the full model 
versus \(\theta\) for three different values of  the mass parameter \(\kappa\) at \(\lambda=1\). 
We show the approach to the continuum limit using \(\beta = 1.6\), \(3.6\), \(6.4\) and \(10.0\) at fixed \(\beta/N_s N_t = 0.1\). 
The discontinuity in the top left plot near \(\theta=\pm 3\pi\) 
is due to the violation of \(\beta>\theta/2\pi\).}
\label{fig3}
\end{figure*}

Now we switch to lighter matter fields to study the continuum limit and to compare the results for three different 
values of the mass parameter, \(\kappa = 4.0\), \(\kappa = 5.0\) and 
again the heavy mass value \(\kappa = 10.0\). As in the case of pure gauge theory we 
consider the continuum limit for \(\beta \longrightarrow \infty\)
with \(\beta/N_s N_t\) fixed, and use \(\beta = 1.6\), \(3.6\), \(6.4\) and \(10.0\) at 
fixed \(\beta/N_s N_t = 0.1\).
In Fig.~\ref{fig3} we show the results for the plaquette expectation value as a function of \(\theta\). As before in the quenched case 
also here we observe that, as we approach the continuum limit, the \(2\pi\)-periodicity of the observable emerges as expected. 
It is remarkable that for this pure gauge observable the dependence  
on the mass parameter \(\kappa\) is only very weak and the behavior of the plaquette  in the full theory is nearly the same as for the 
pure gauge case as long as we are in the symmetric phase.

\begin{figure*}[t]
\begin{center}
\includegraphics[height=6.6cm,type=pdf,ext=.pdf,read=.pdf]{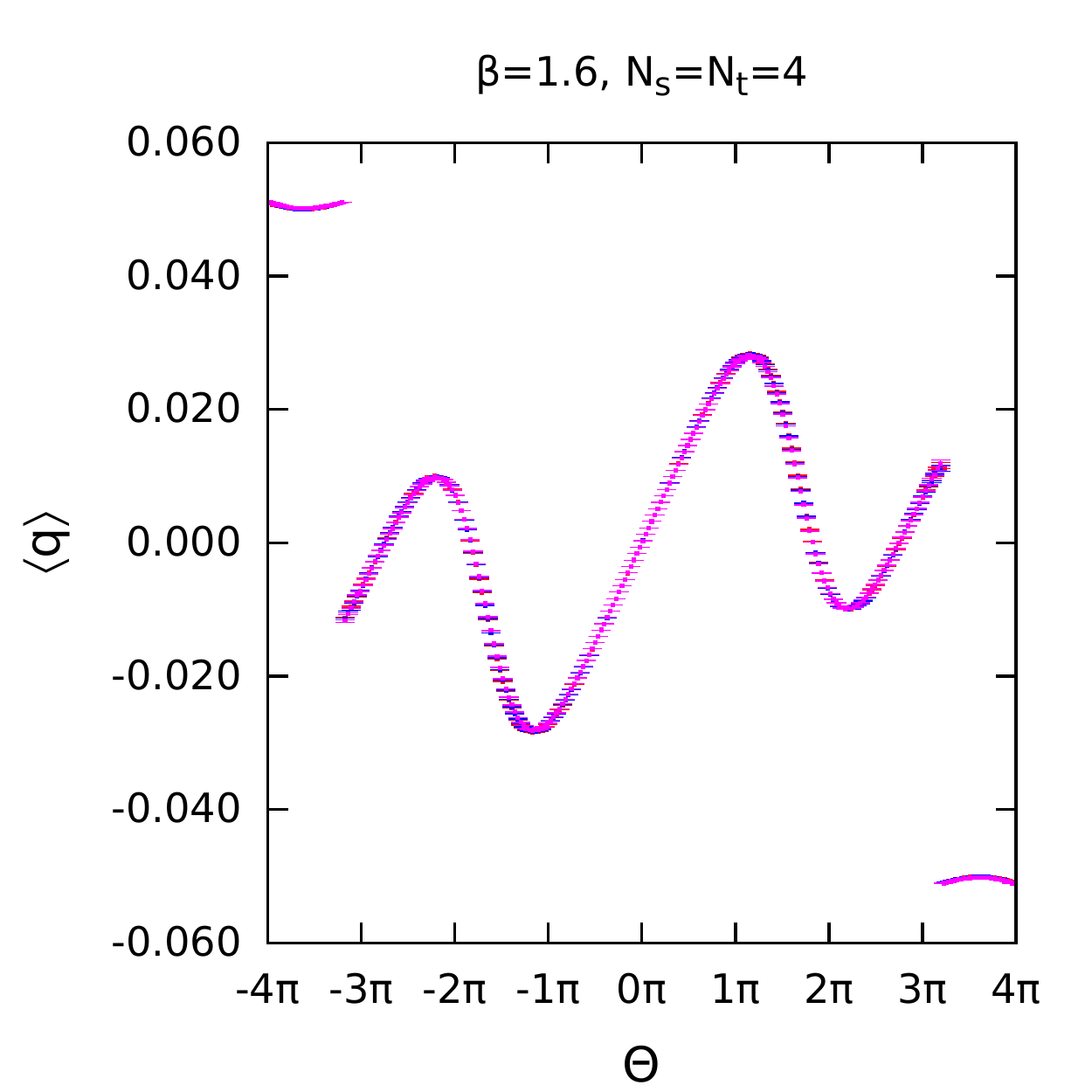}
\hskip8mm
\includegraphics[height=6.6cm,type=pdf,ext=.pdf,read=.pdf]{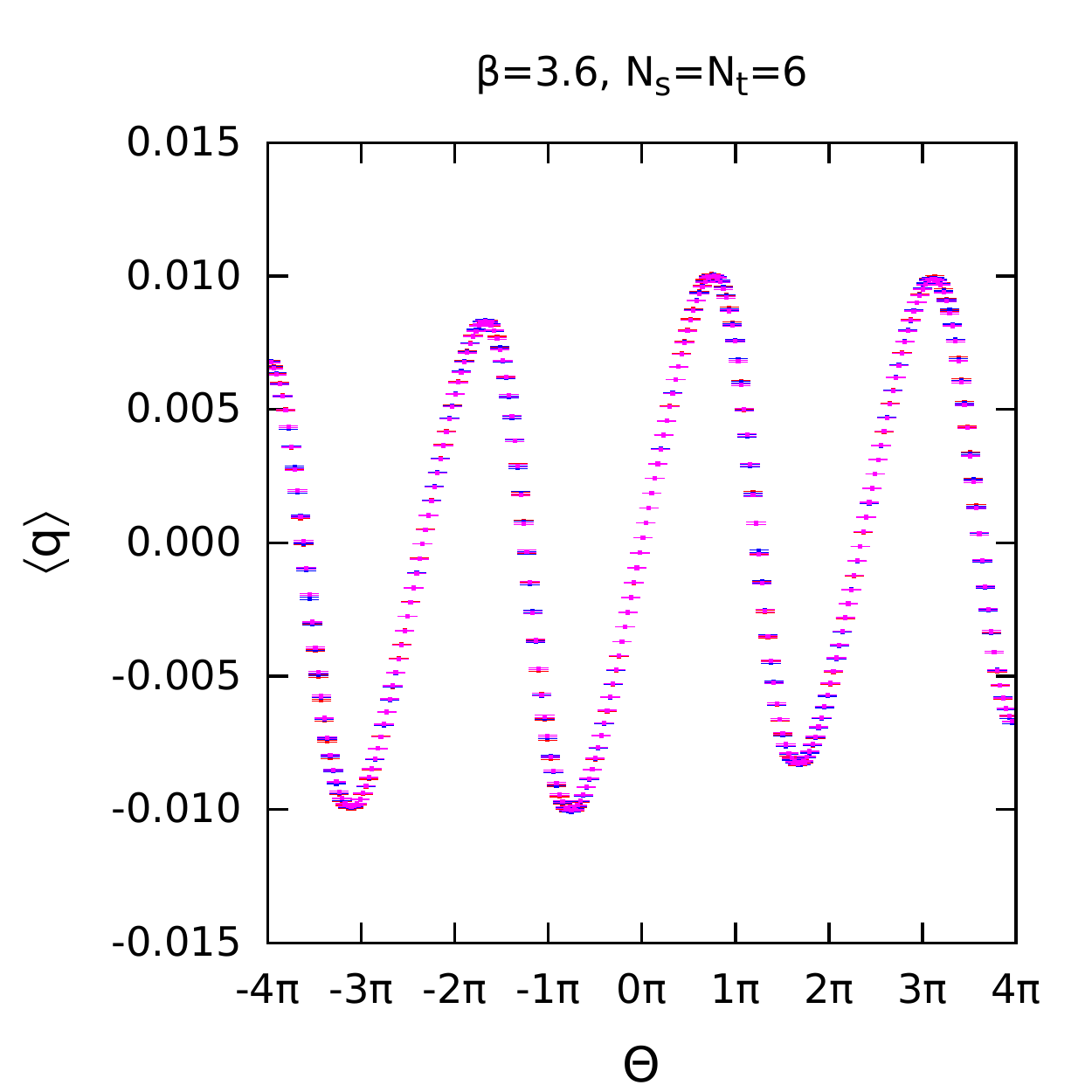}

\includegraphics[width=6.6cm,type=pdf,ext=.pdf,read=.pdf]{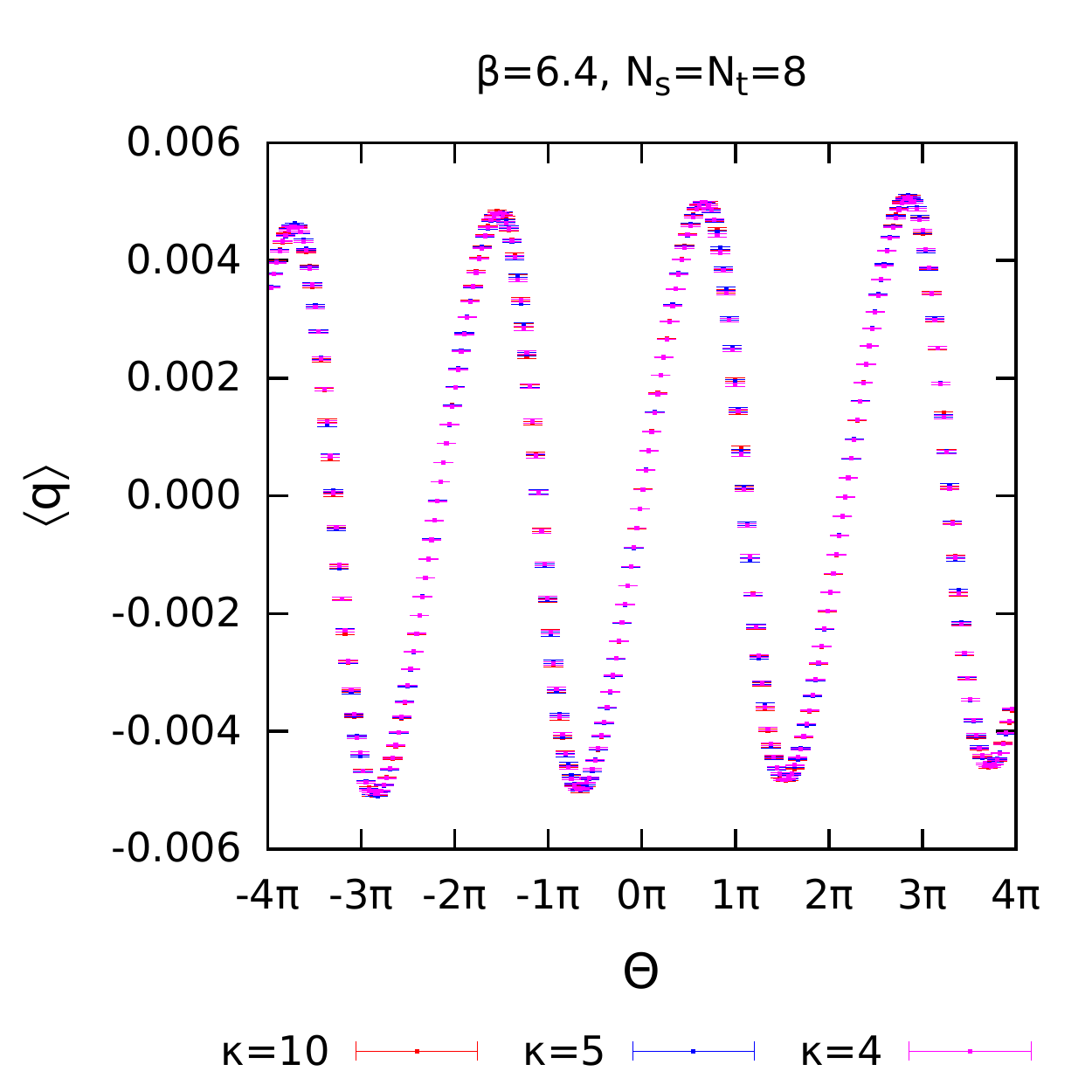}
\hskip8mm
\includegraphics[width=6.6cm,type=pdf,ext=.pdf,read=.pdf]{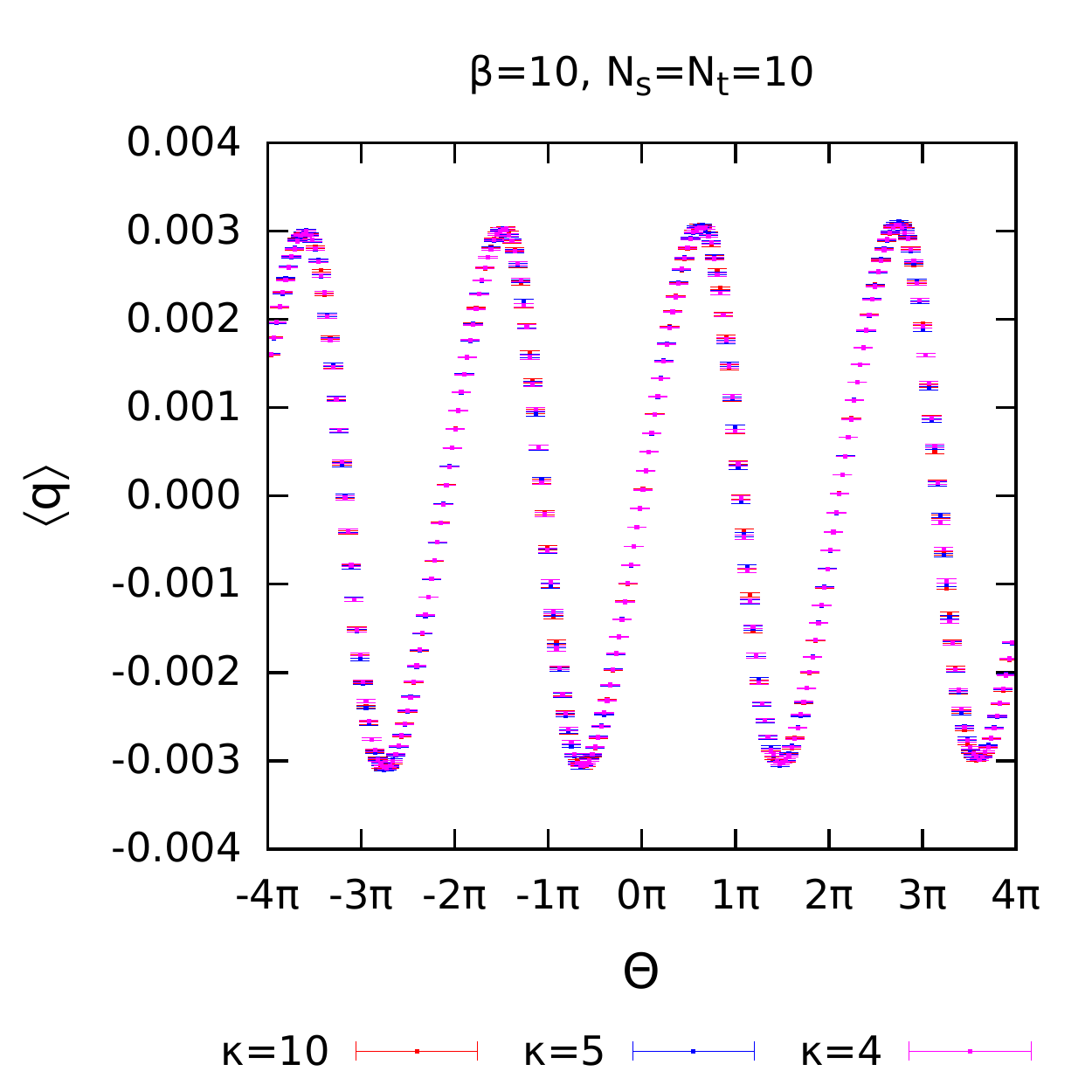}
\end{center}
\caption{The topological charge density $\langle q \rangle$ versus the vacuum angle \(\theta\) 
for three different values of  the mass parameter \(\kappa\) at \(\lambda=1\).
We show the approach to the continuum limit using \(\beta = 1.6, 3.6, 6.4\) and \(10.0\) at fixed $\beta/N_s N_t = 0.1$. 
Again the discontinuity in the top left plot near \(\theta=\pm 3\pi\) 
is due to the violation of \(\beta>\theta/2\pi\).}
\label{fig4}
\end{figure*}

Finally, in Fig.~\ref{fig4} we study the expectation value of the topological charge density
$\langle q \rangle$ as a function of \(\theta\), again using the values \(\kappa = 4.0\), \(5.0\) and \(10.0\) and 
approaching the continuum limit with the sequence \(\beta = 1.6\), \(3.6\), \(6.4\) and \(10.0\) at fixed \(\beta/N_s N_t = 0.1\). 
As before we observe the emergence of $2\pi$-periodicity in the continuum limit. 

Here we remark that the topological charge density  $\langle q \rangle$ is a variable that is odd in \(\theta\), while the plaquette 
$\langle \mbox{Re} \,U_p \rangle$ is an even function. This can be seen from the fact that when changing the link variables in the
path integral according to  \(U_{x,p} \rightarrow U_{x,p}^{\,*}\) we have $Q[U] \rightarrow - Q[U]$, while
$ \mbox{Re} \, U_p $ remains unchanged (compare Eqs. (\ref{zconventional}) and (\ref{topcharge})).
This leads to a linearly rising behavior of $\langle q \rangle$ in the vicinity of \(\theta = 0\). This (anti-) symmetry 
is clearly visible in all plots of Fig.~\ref{fig4}, while the expected \(2\pi\)-periodicity in \(\theta\) is recovered only in the continuum limit.

As already for the plaquette expectation value, also for the topological charge density $\langle q \rangle$ we find 
that the results are essentially independent of the mass parameter $\kappa$. This indicates that the physics of the model 
related to the vacuum angle $\theta$ is dominated by the behavior of the gauge sector and the dynamics of the 
matter field plays only a sub-leading role as long as we are in the symmetric phase (c.f. Sec. \ref{sec_phase_diagram}). 
As we will see below, the behavior changes considerably in the Higgs phase.

\subsection{Phase diagram in the $\kappa$-$\lambda$ plane}
\label{sec_phase_diagram}

We now discuss the determination of the phase diagram in the plane of the mass parameter \(\kappa\) and the quartic coupling \(\lambda\). 
In more than two dimensions one expects a true phase transition line separating the symmetric from the Higgs phase. In two 
dimensions one only finds smooth behavior of the variables since spontaneous breaking of a continuous symmetry is not possible in 
two dimensions due to the Mermin-Wagner-Coleman theorem. Second order derivatives of the free energy show an extremum but no 
scaling with the volume. 

\begin{figure*}[t]
\begin{center}
\includegraphics[height=7cm,type=pdf,ext=.pdf,read=.pdf]{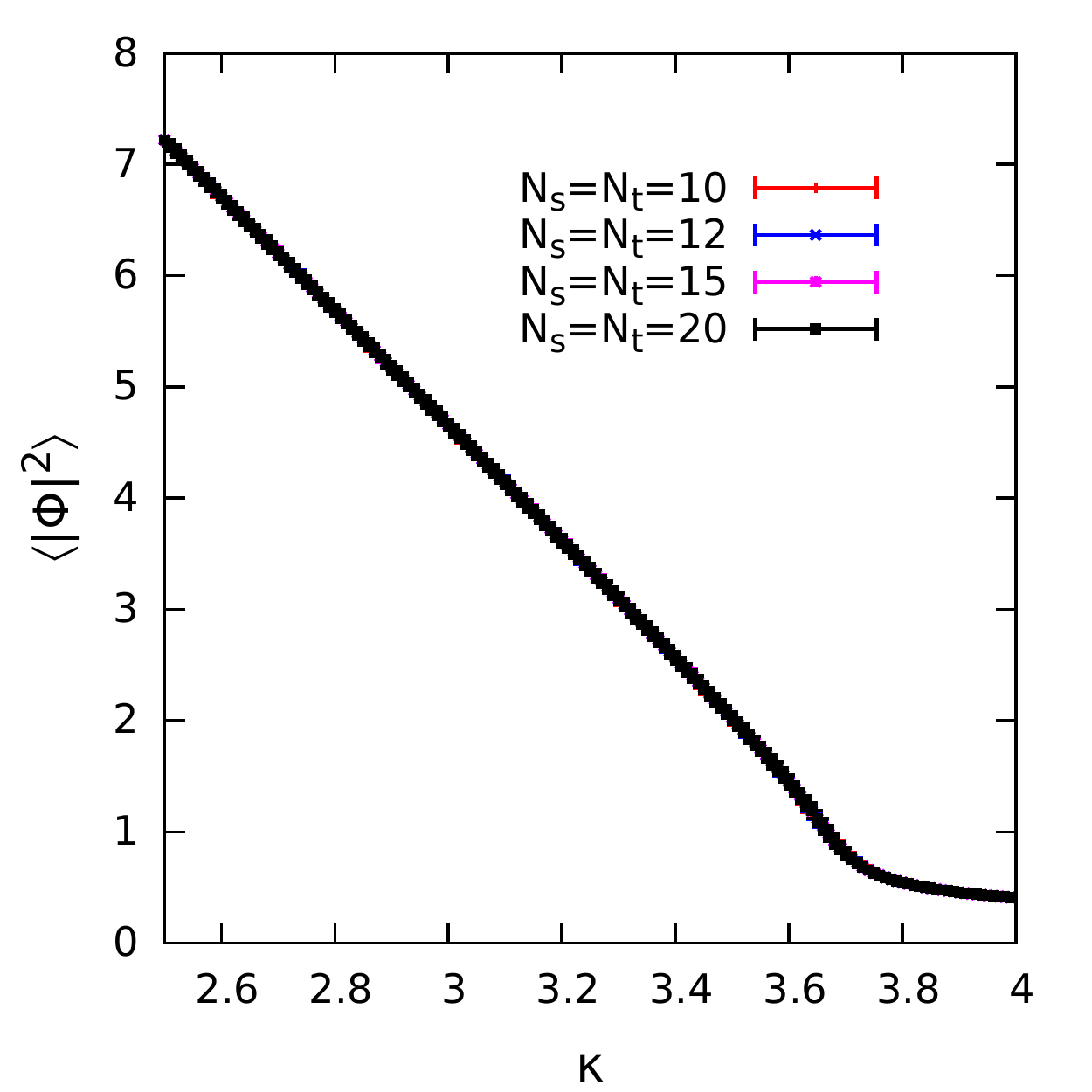}
\hskip10mm
\includegraphics[height=7cm,type=pdf,ext=.pdf,read=.pdf]{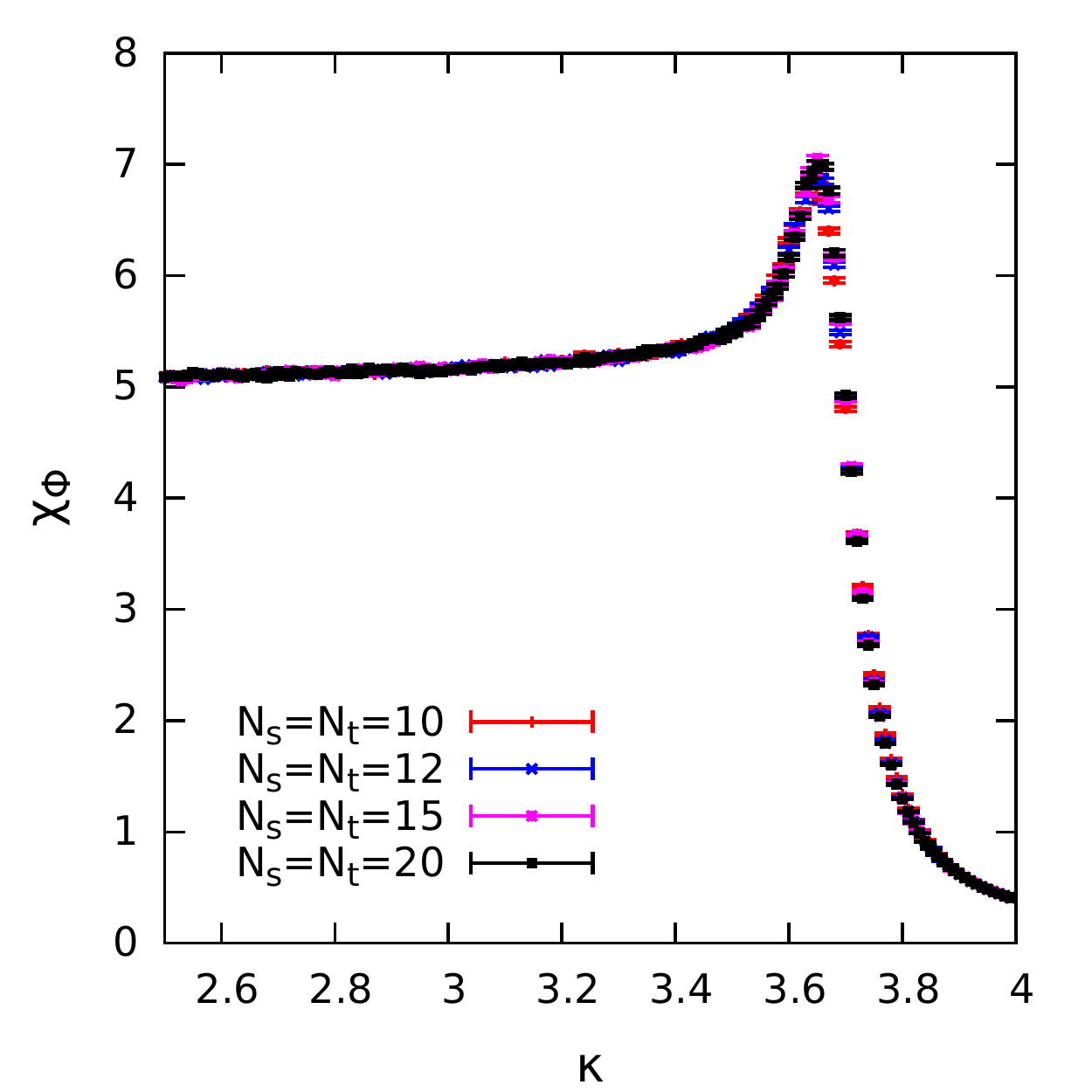}
\end{center}
\caption{Expectation value of the square of the field modulus $\langle |\phi|^2 \rangle$ and the corresponding susceptibility
\(\chi_{\phi}\) versus the mass parameter \(\kappa\) at \(\lambda=0.1\) and \(\beta=10.0\) for different volumes. One can clearly 
see the changing behavior of the first derivative starting at approximately \(\kappa=3.7\), leading to a peak in the corresponding 
susceptibility. The observable is essentially independent of the volume indicating a smooth transition.}
\label{fig_absphi2}
\end{figure*}

An example is given in Fig.~\ref{fig_absphi2} where we show the field expectation value $\langle |\phi|^2 \rangle$ (lhs.~plot) and the
corresponding susceptibility \(\chi_{\phi}\) (rhs.) versus the mass parameter \(\kappa\) at \(\lambda=0.1\) and \(\beta=10.0\) for different 
volumes. It is obvious that the data for different volumes fall on top of each other and volume scaling is absent.

\begin{figure}[t]
\begin{center}
\includegraphics[height=6cm,type=pdf,ext=.pdf,read=.pdf]{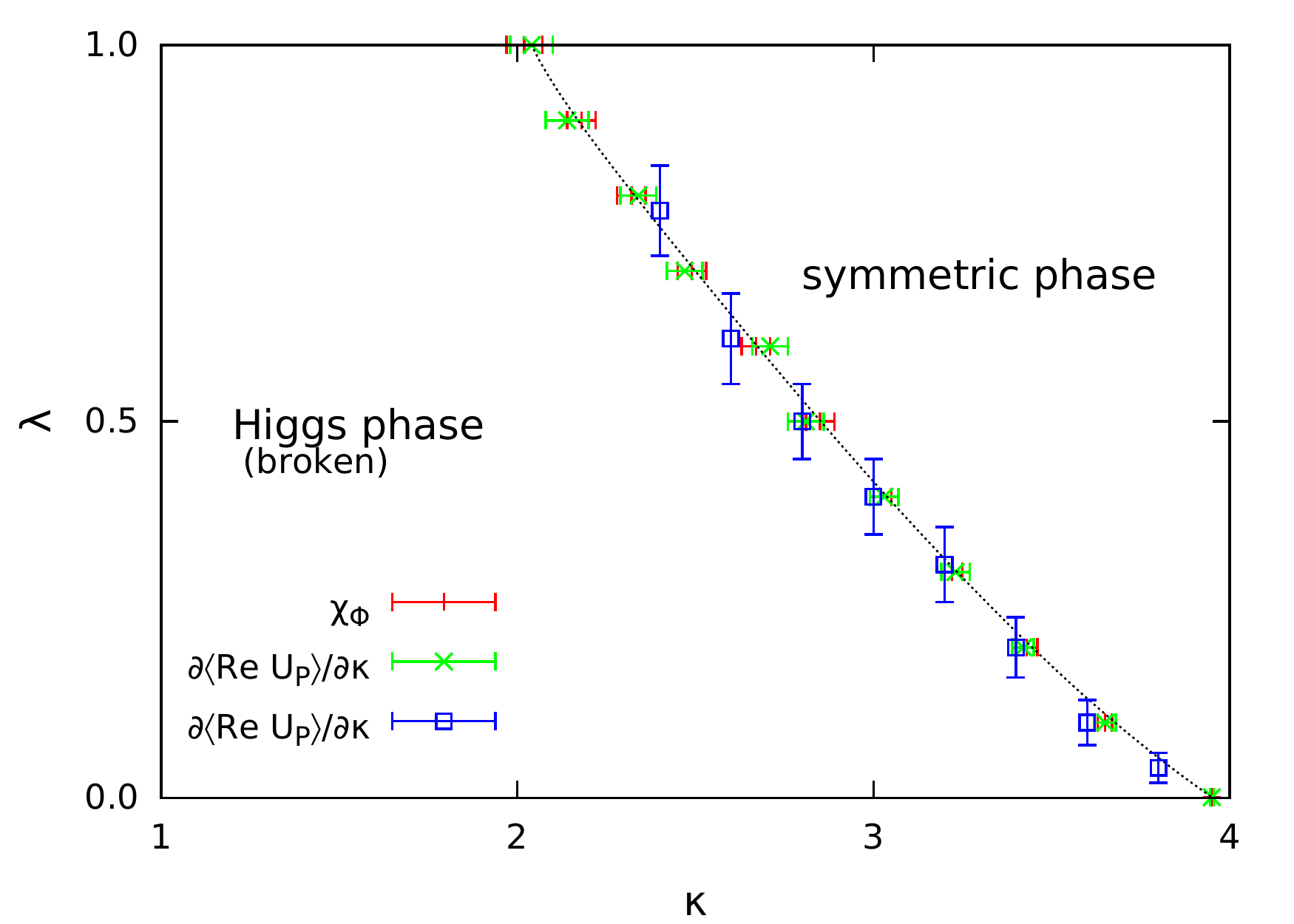}
\end{center}
\caption{Phase diagram in the plane of couplings \(\lambda\) and \(\kappa\). We find two phases separated by a crossover line.
The phase boundary was determined from the maxima of \(\chi_{\phi}\) as a function of $\kappa$ (plusses connected with a dotted line) 
and the maxima 
of $\partial / \partial \kappa \langle \mbox{Re} \, U_p \rangle$ as a function of both, $\kappa$ (crosses) and of $\lambda$ (squares).}
\label{fig_kl_phase_diagram}
\end{figure}

Nevertheless, one expects physical properties to change when crossing the line of the smooth transition and in order to determine 
the location of this line in the $\kappa$-$\lambda$ plane, i.e., the phase diagram, we identify the position of the peak in higher 
derivatives for various values of the parameters. The results are shown in Fig.~\ref{fig_kl_phase_diagram}.
For the determination of the transition line in Fig.~\ref{fig_kl_phase_diagram} we used the maxima of \(\chi_{\phi}\) as a function 
of $\kappa$ (plusses) and the maxima of $\partial / \partial \kappa \langle \mbox{Re} \, U_p \rangle$ as a function of both, 
$\kappa$ (crosses) and of $\lambda$ (squares).
We stress, that at a crossover transition different higher order derivatives do not necessarily have to peak at the same position. 
However, as is obvious from Fig.~\ref{fig_kl_phase_diagram}, we here find
agreement among all observables within error bars.  We find a line of crossover points extending from \(\kappa=4.0\) 
and \(\lambda=0\) (which corresponds to the free massless field) to \(\kappa=2.0\) and \(\lambda=1\) (the largest value of $\lambda$ 
considered here). The data for Fig.~\ref{fig_kl_phase_diagram} are for \(N_s=N_t=10\) and $\beta = 10.0$, i.e., relatively close to 
the continuum limit.

\begin{figure}[t]
\begin{center}
\includegraphics[height=7.5cm,type=pdf,ext=.pdf,read=.pdf]{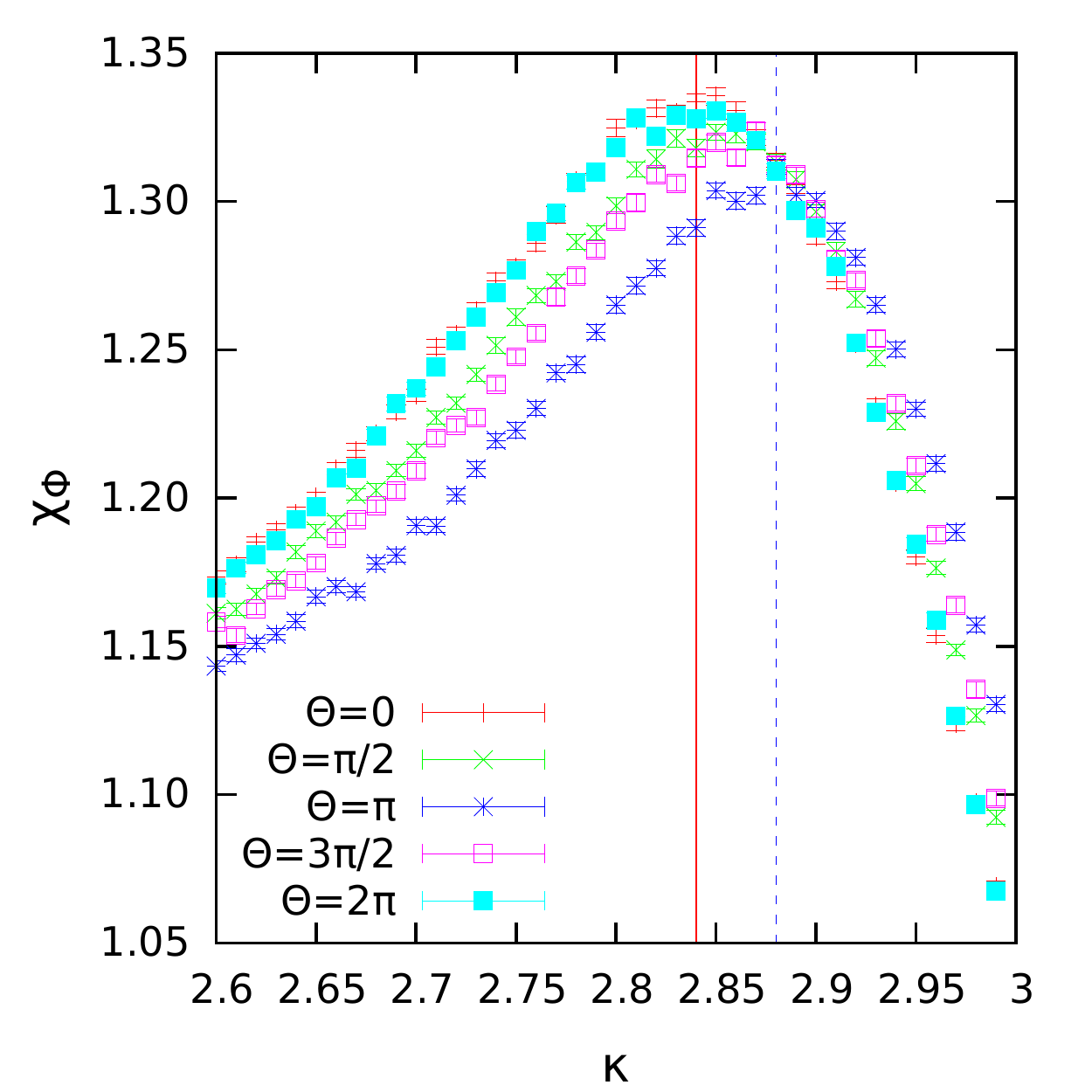}
\end{center}
\caption{Variation of the peak in \(\chi_{\phi}\) with \(\theta\) at \(\lambda=0.5\), \(\beta=10.0\) and \(N_s=N_t=10\). The vertical lines 
delimit the range of $\kappa$-values where the cross-over is found (as determined from the maxima of $\chi_{\phi}$)
when varying $\theta$ between \(\theta=0\) (full red line) to \(\theta=\pi\) (dashed blue line) and back to \(\theta=2\pi\).}
\label{fig_chiabsphi_vs_k_t}
\end{figure}

An interesting question is how strong the position of the crossover line depends on the vacuum angle $\theta$. In our study we find 
that this dependence is only very weak. This is illustrated in Fig.~\ref{fig_chiabsphi_vs_k_t}, where the peak of \(\chi_{|\phi|^2}\) is shown 
in high resolution for four different values of \(\theta\in[0,2\pi]\) at \(\lambda=0.5\). Starting at \(\theta=0\) the peak gets shifted towards 
slightly larger values of the mass parameter (here from \(\kappa\approx2.84\) to \(\kappa\approx2.88\)) until \(\theta=\pi\), 
after which, due to periodicity in $\theta$, the peak moves back to its position at \(\theta=0\) when we reach \(\theta=2\pi\). 
The situation is similar for other values of the coupling \(\lambda\in[0,1]\). We conclude that the variation of \(\theta\) amounts to only 
a very small shift of the crossover line.

\subsection{Characterization of the two phases}
\label{sec_phase_characterization}

\begin{figure*}[t]
\begin{center}
\includegraphics[height=6cm,type=pdf,ext=.pdf,read=.pdf]{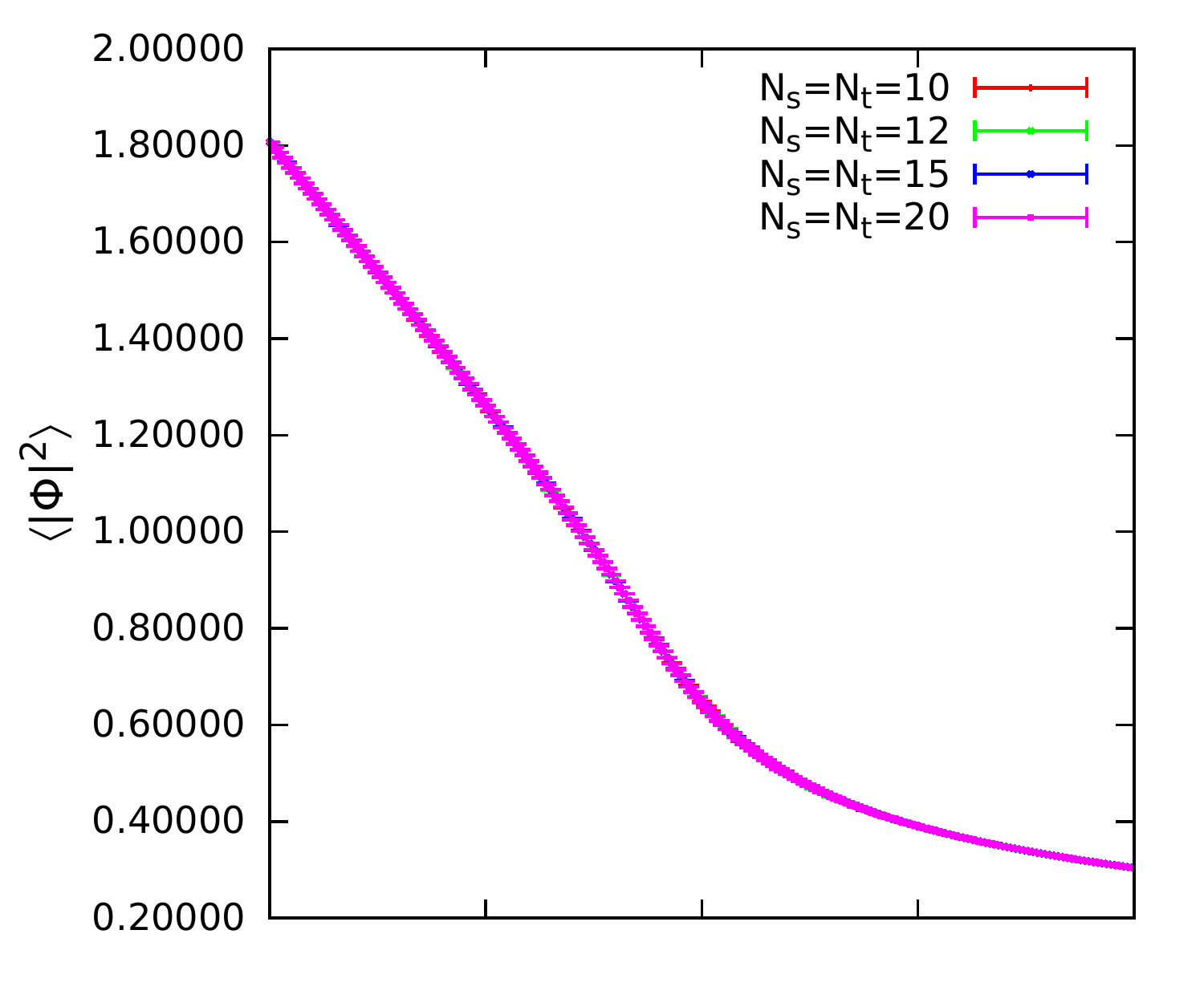}
\includegraphics[height=6cm,type=pdf,ext=.pdf,read=.pdf]{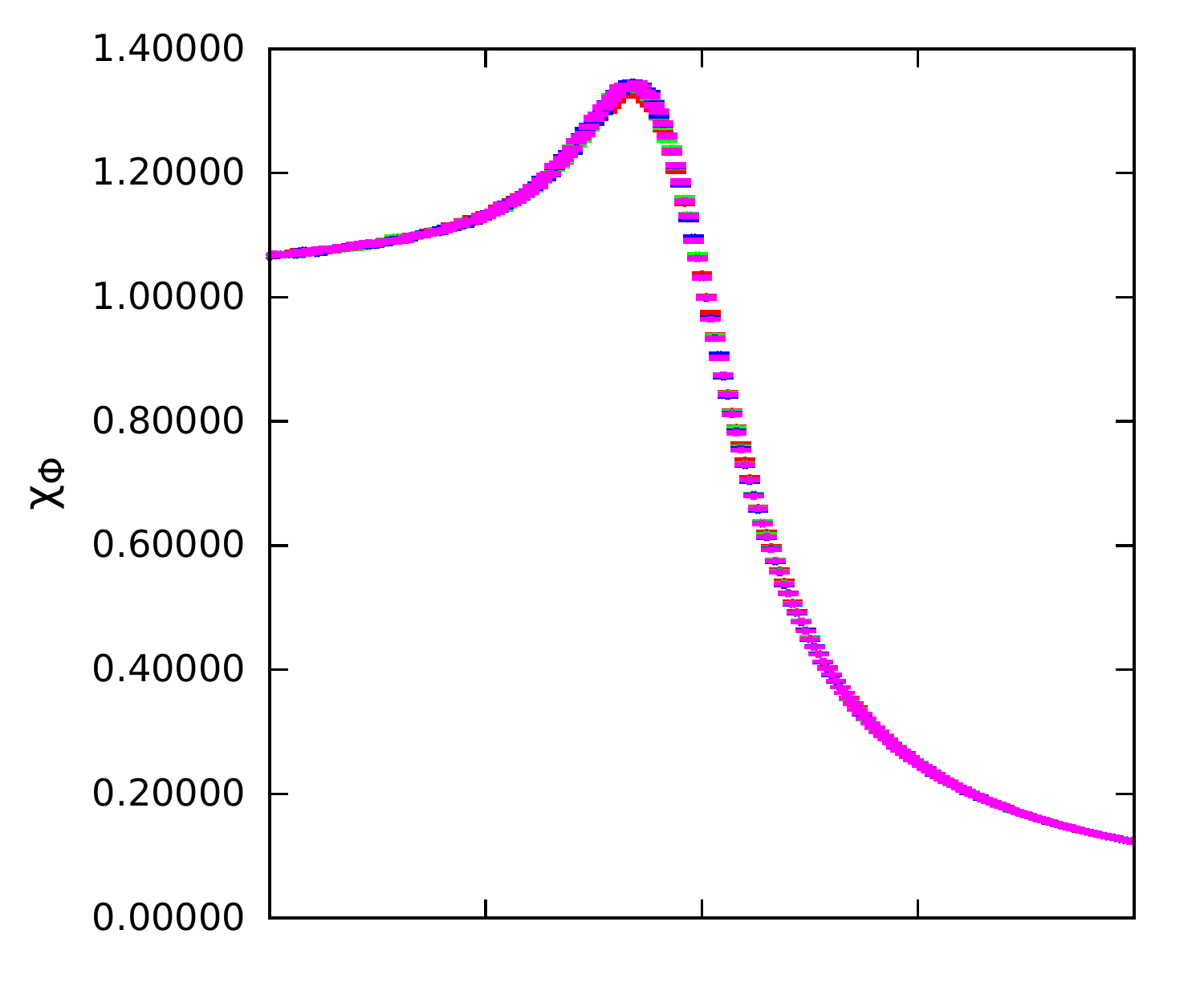}

\vspace*{-6mm}

\includegraphics[height=6cm,type=pdf,ext=.pdf,read=.pdf]{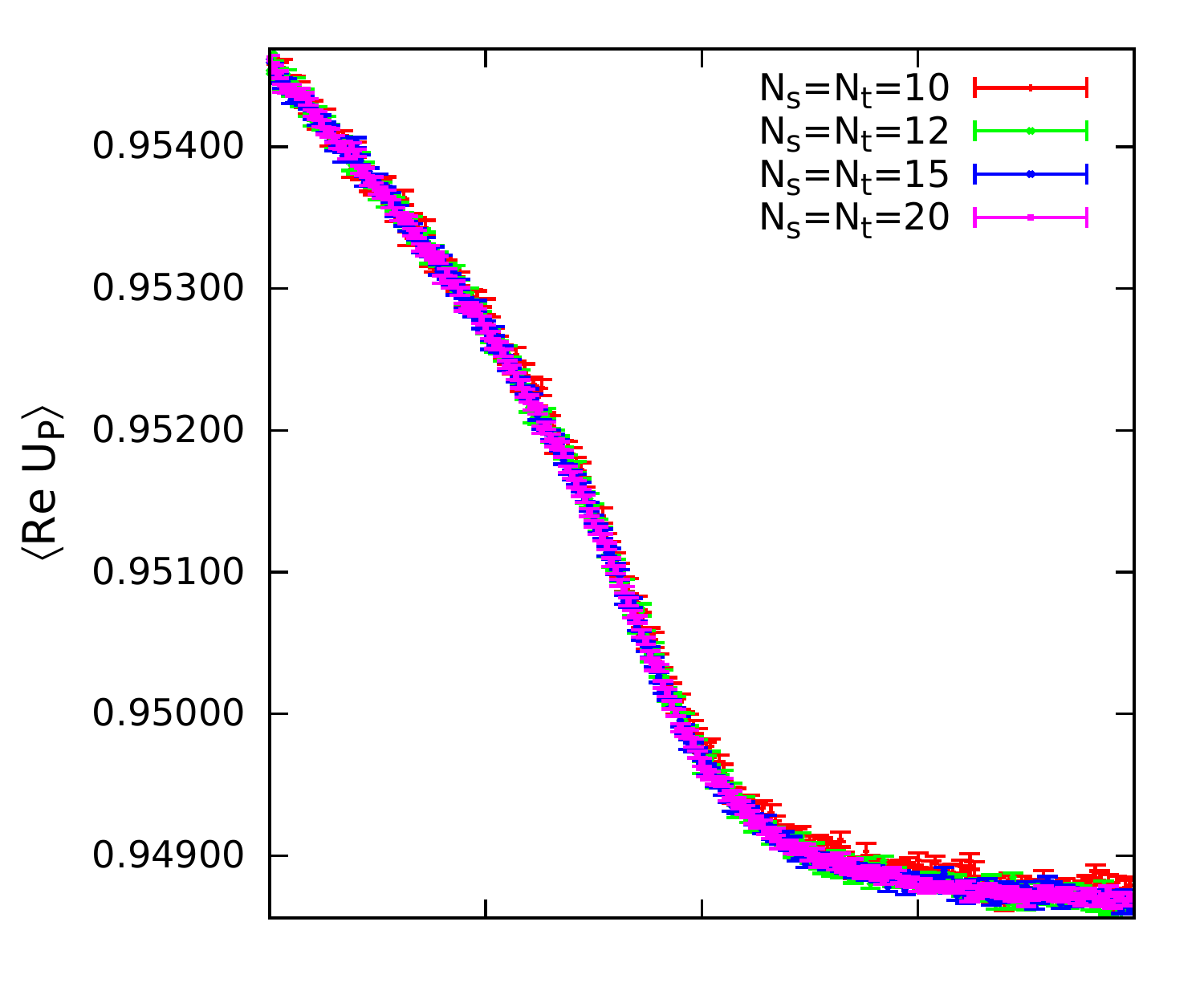}
\includegraphics[height=6cm,type=pdf,ext=.pdf,read=.pdf]{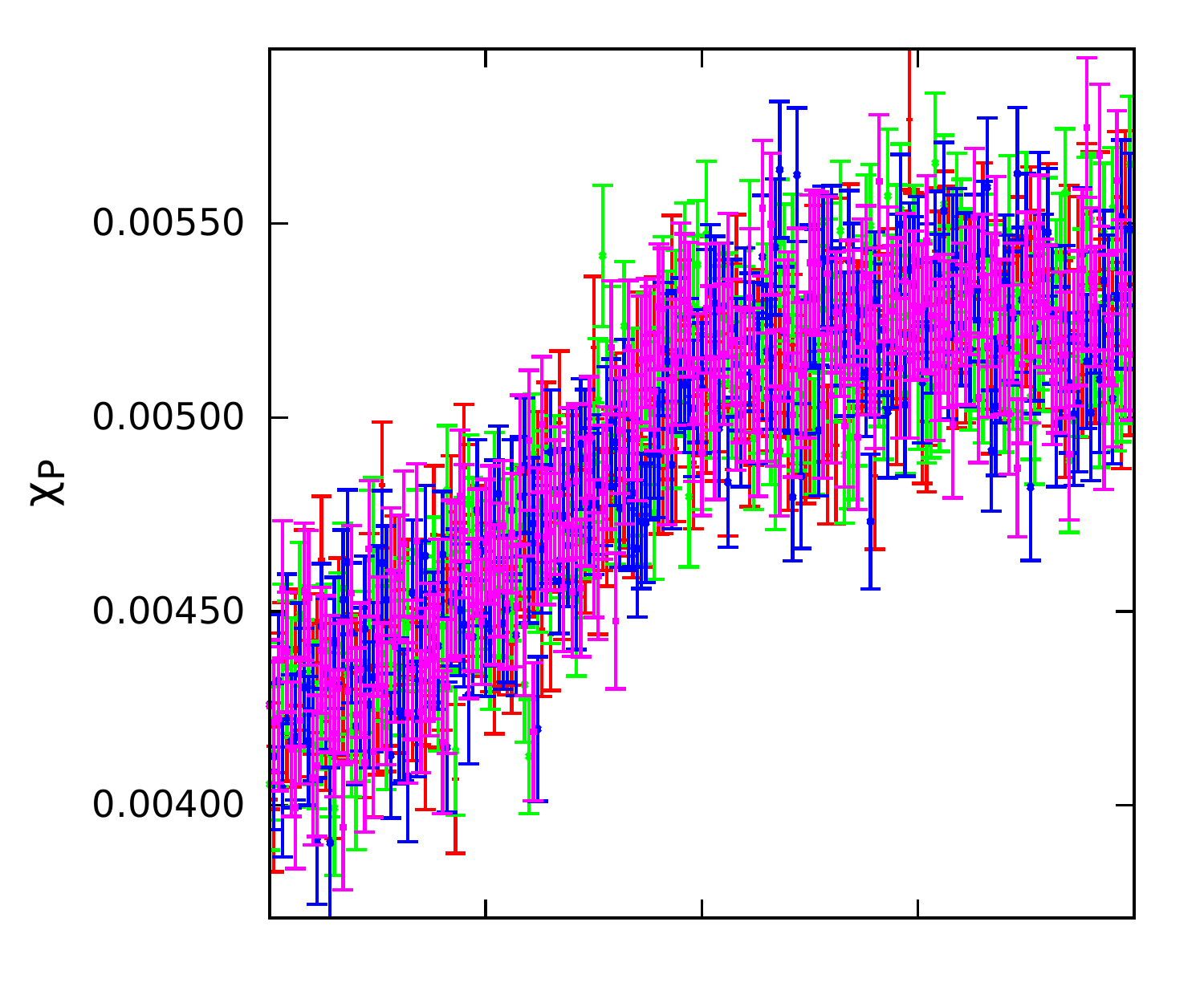}

\vspace*{-6mm}

\includegraphics[height=6cm,type=pdf,ext=.pdf,read=.pdf]{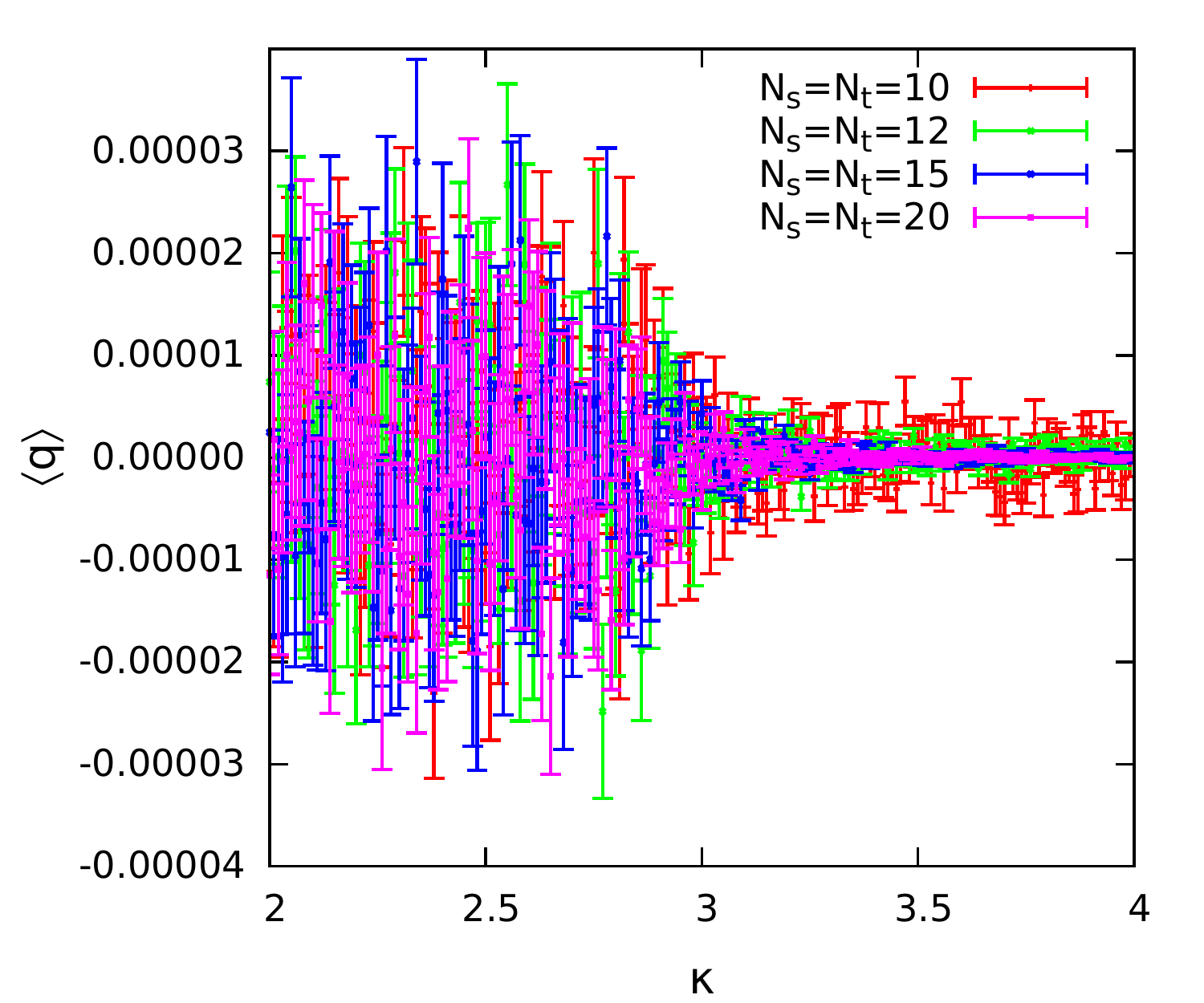}
\includegraphics[height=6cm,type=pdf,ext=.pdf,read=.pdf]{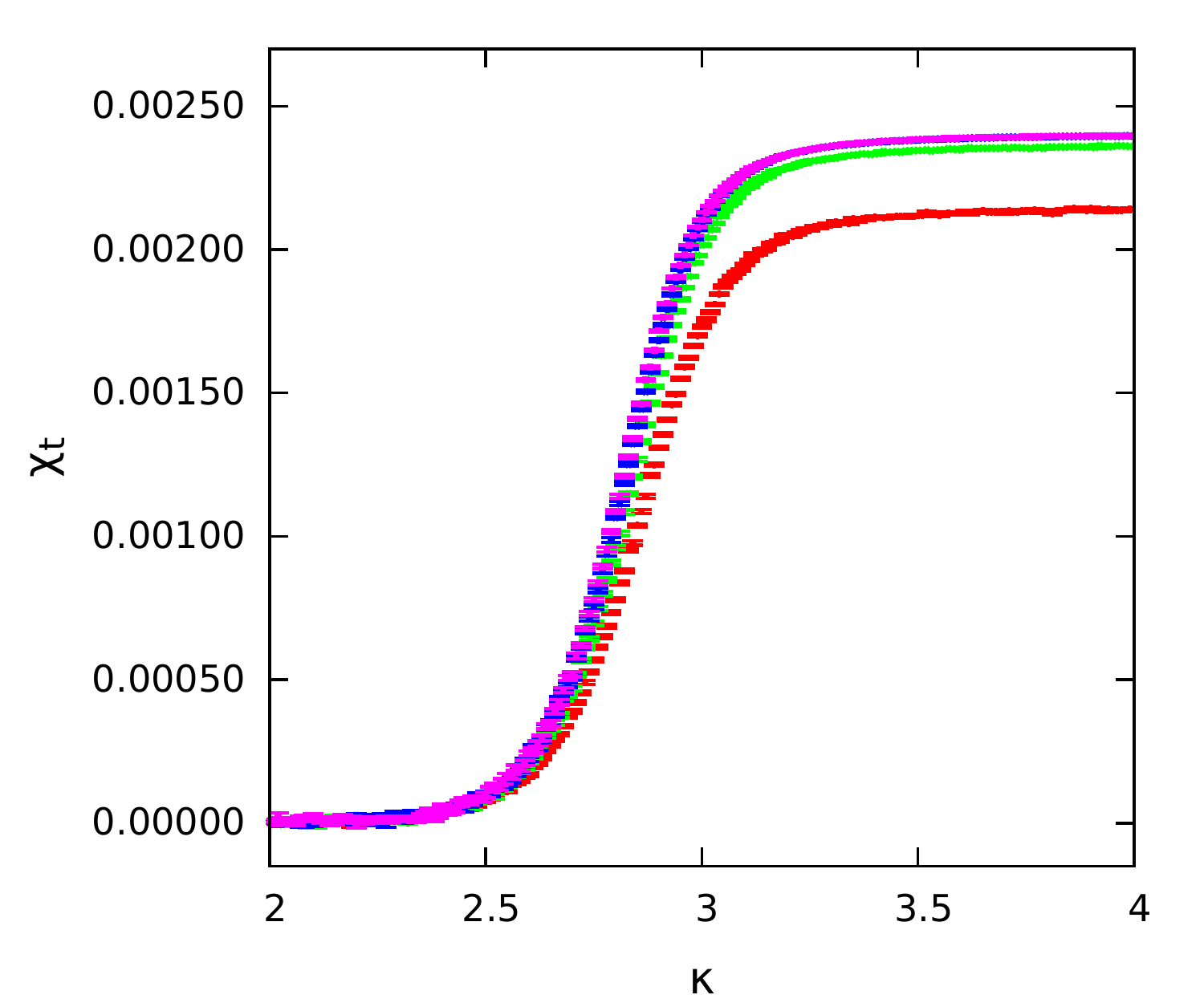}
\end{center}
\caption{Observables versus the mass parameter \(\kappa\) at \(\lambda=0.5\), \(\beta=10.0\) 
and \(\theta=0\) for different volumes. One can clearly see the changing behavior of all observables 
at approximately \(\kappa=2.8\) corresponding to the crossover position at $\lambda = 0.5$ as mapped out 
in the phase diagram in the previous section.}
\label{fig_obs_vs_k}
\end{figure*}

As a next step in our explorative study of using dual variables in the U(1) gauge-Higgs model with topological term we attempt a characterization 
of the two phases - in particular with respect to their dependence on the vacuum angle $\theta$. We remark again that  the transition is 
smooth (it is of the Kosterlitz-Thouless type) and no simple order parameter exists. Nevertheless, for completeness, we start with 
summarizing the characteristic behavior of all observables in the two phases for $\theta = 0$. The corresponding results are presented 
in Fig. \ref{fig_obs_vs_k}, where the six observables discussed above are shown as a function of the mass parameter 
\(\kappa\) for $\lambda = 0.5$. In other words, we analyze horizontal slices through the phase diagram Fig.~\ref{fig_kl_phase_diagram} 
and we compare the results for four different volumes. In all observables a changing behavior can be seen near $\kappa = 2.8$, the
value on the critical line for $\lambda = 0.5$.  The topological charge is (within error bars) identically zero, as expected at \(\theta=0\),
but obviously the fluctuations are different in the two phases, as can be seen by the change of the size of the error bars and the increase of the 
fluctuations. The observables do not depend on the lattice volume, except for \(\chi_{top}\) which shows a 
small finite volume effect in the symmetric phase for the smallest lattice size (see also the discussion below). 
Let us notice at this point that the behavior of the topological susceptibility versus $\kappa$
for fixed $\lambda$ and $\beta$ at $\theta=0$ has been checked by simulating the model with
the standard Metropolis algorithm without dualization of the variables. We found exact 
agreement.  

At this point we remark that the topological susceptibility at $\theta \ne 0$
can also be negative, despite the fact that it is a second derivative 
of $\ln Z$. This can be seen as follows: Instead of integrating in the path integral over the variables $U_{x,\mu}, \phi_x$ we can also integrate 
over the complex conjugate variables $U_{x,\mu}^{\; *}, \phi_x^{\, *}$. 
Using the symmetry properties (compare equations (\ref{zconventional}) - (\ref{topcharge}))
$\int {\cal D}[U^*] = \int {\cal D}[U]$,  $\int {\cal D}[\phi^*] = \int {\cal D}[\phi]$, 
$S_G[U^*] = S_G[U]$, $S_M[U^*,\phi^*] = S_M[U,\phi]$ and $Q[U^*] = - Q[U]$ we find for the partition sum
\begin{align}
Z  \; &=  \; \int \! \mathcal{D}[U]\mathcal{D}[\phi] \; e^{-\, S_G[U] \, - \, S_M[U,\phi]} \; e^{ -i\,\theta \, Q[U]}\\
&=  \; 
\int \! \mathcal{D}[U]\mathcal{D}[\phi] \; e^{-\, S_G[U] \, - \, S_M[U,\phi]} \; \cos( \theta \, Q[U] ) \; . \nonumber 
\end{align}
Evaluating the topological susceptibility as the second derivative of $\ln Z$ with respect to $\theta$ (compare Eq.~(\ref{topobs})) we obtain
\begin{align}
&N_s\,N_t~\chi_{t} 
= -\frac{\partial^2}{\partial \theta^2} \ln Z \\
& = \frac{1}{Z} 
\int \! \mathcal{D}[U]\mathcal{D}[\phi] \; e^{-\, S_G[U] - S_M[U,\phi]} \cos( \theta \, Q[U] ) \; Q[U]^2 
\nonumber \\
&+ \!
\left( \frac{1}{Z} \!\int \! \mathcal{D}[U]\mathcal{D}[\phi] e^{-\, S_G[U] - S_M[U,\phi]} \sin( \theta \, Q[U] ) Q[U]\right)^2 \!\!. \nonumber
\end{align}
The right hand side has two contributions, with the second term being the square of the first derivative and thus positive. The first term,
however, comes with $\cos( \theta \, Q[U] )$, i.e., the even part of $\exp( - i \theta \, Q[U] )$. This first term can be negative, and thus also 
$\chi_{t}$ needs not be positive. The same of course also holds for the dual expression of $\chi_{t}$ given in (\ref{dualobs2}).

\begin{figure*}[t]
\begin{center}
\hspace*{-5mm}
\includegraphics[height=7.3cm,type=pdf,ext=.pdf,read=.pdf]{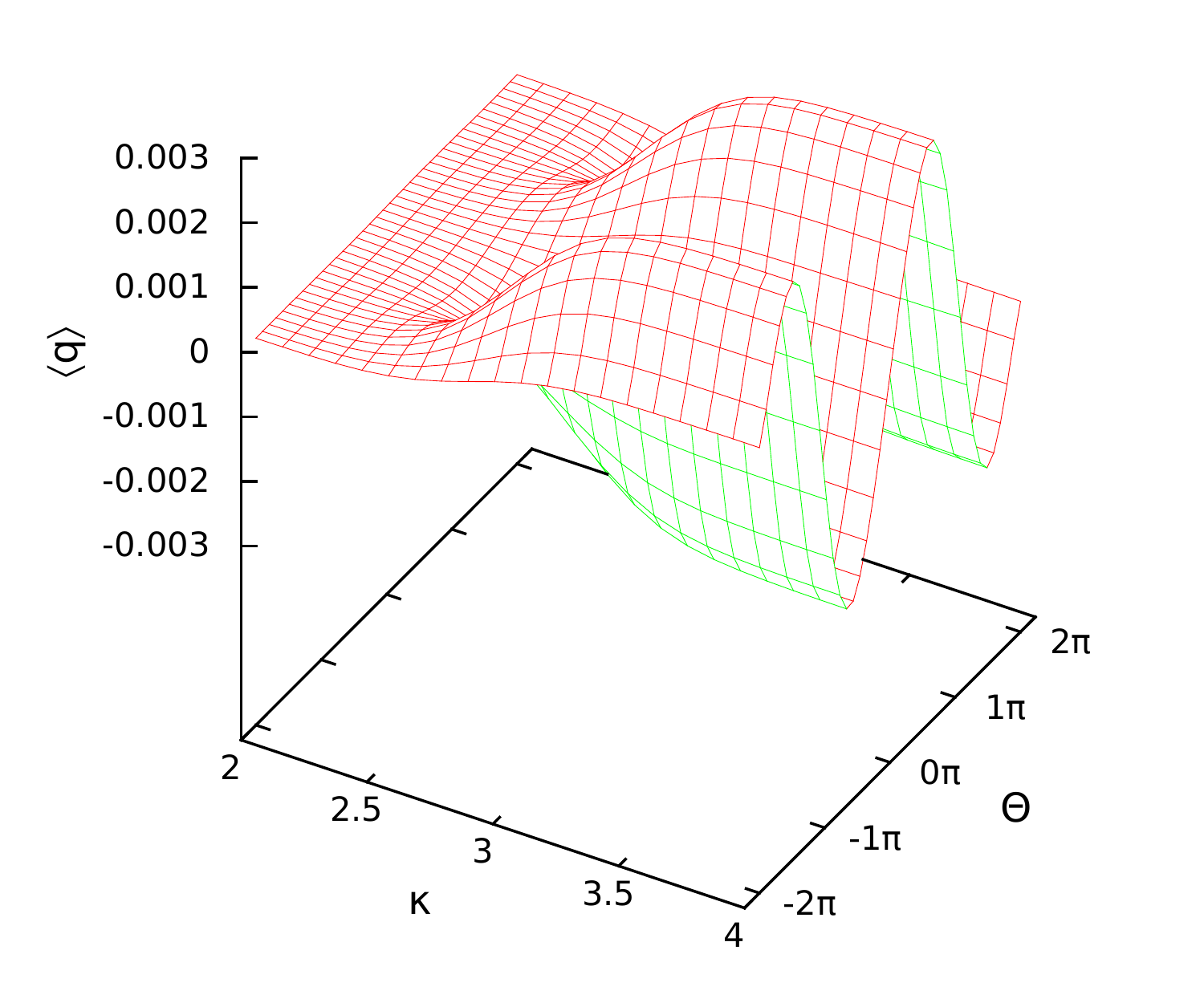}
\hspace*{-8mm}
\includegraphics[height=7.3cm,type=pdf,ext=.pdf,read=.pdf]{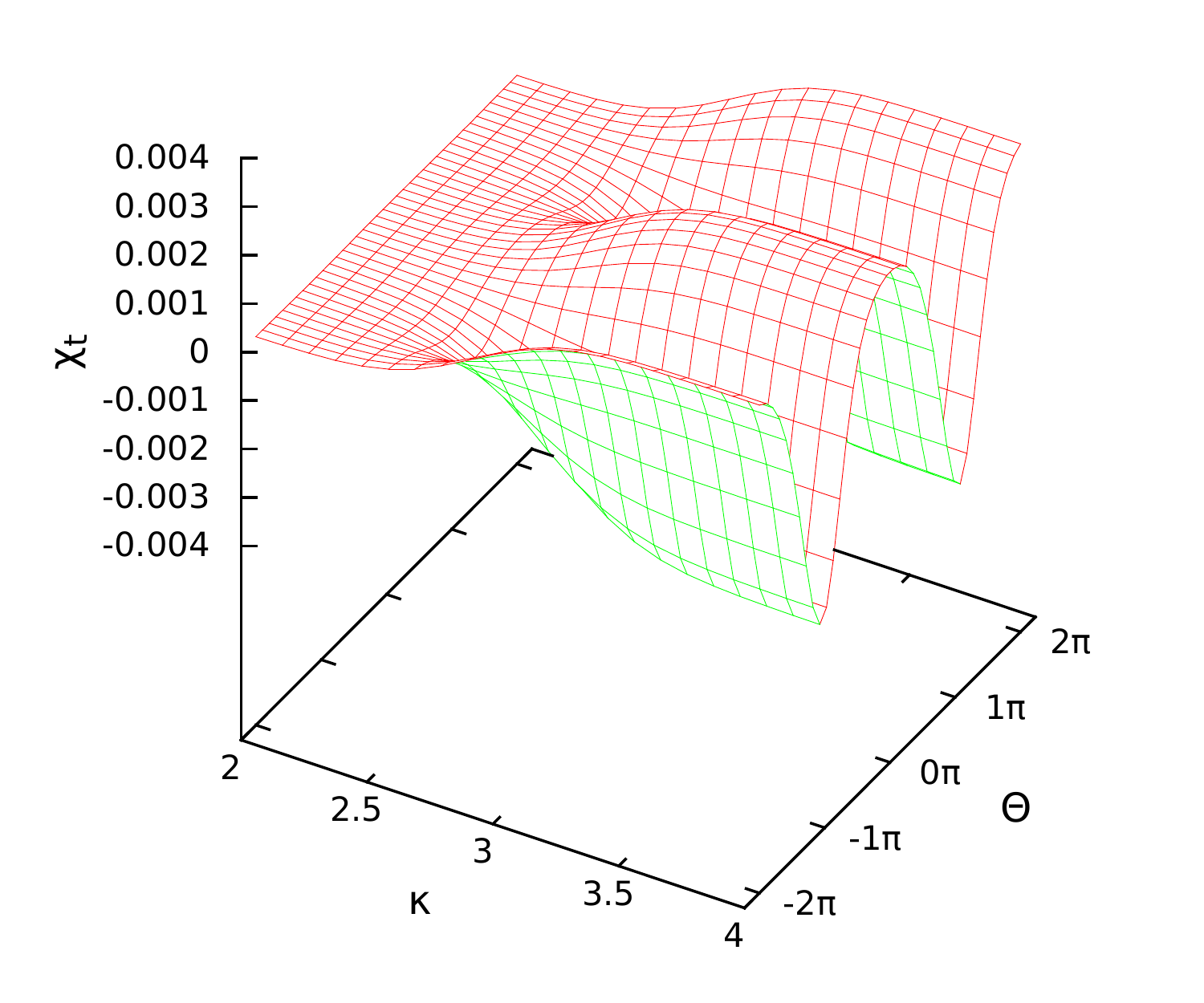}
\end{center}
\caption{Topological charge density (lhs.~plot) and topological susceptibility (rhs.) versus 
\(\kappa\) and \(\theta\) at \(\lambda=0.5\), \(\beta=10\) and \(N_s=N_t=10\).}
\label{fig_3d}
\end{figure*}

In our analysis of the model with the dual variables approach we find that the two phases can be characterized by their response to 
the variation of the vacuum angle. As a first illustration of this fact in Fig. \ref{fig_3d} we show 3-D plots of \(Q\) and \(\chi_{t}\) as a function 
of \(\kappa\) and \(\theta\) at $\lambda = 0.5$, i.e., when considered as a function of $\kappa$ we again inspect a horizontal 
slice through the phase diagram Fig.~\ref{fig_kl_phase_diagram} and we expect to see changing behavior at $\kappa \sim 2.8$. This is indeed
what we observe:  At large values of the mass parameter $\kappa$, i.e., in the symmetric phase, we see oscillatory behavior with $\theta$ 
in both observables, while in the crossover region the observables are independent of $\theta$ within error bars. The transition between the 
two types of behavior takes place as expected near $\kappa \sim 2.8$.

\begin{figure*}[t]
\begin{center}
\includegraphics[height=6cm,type=pdf,ext=.pdf,read=.pdf]{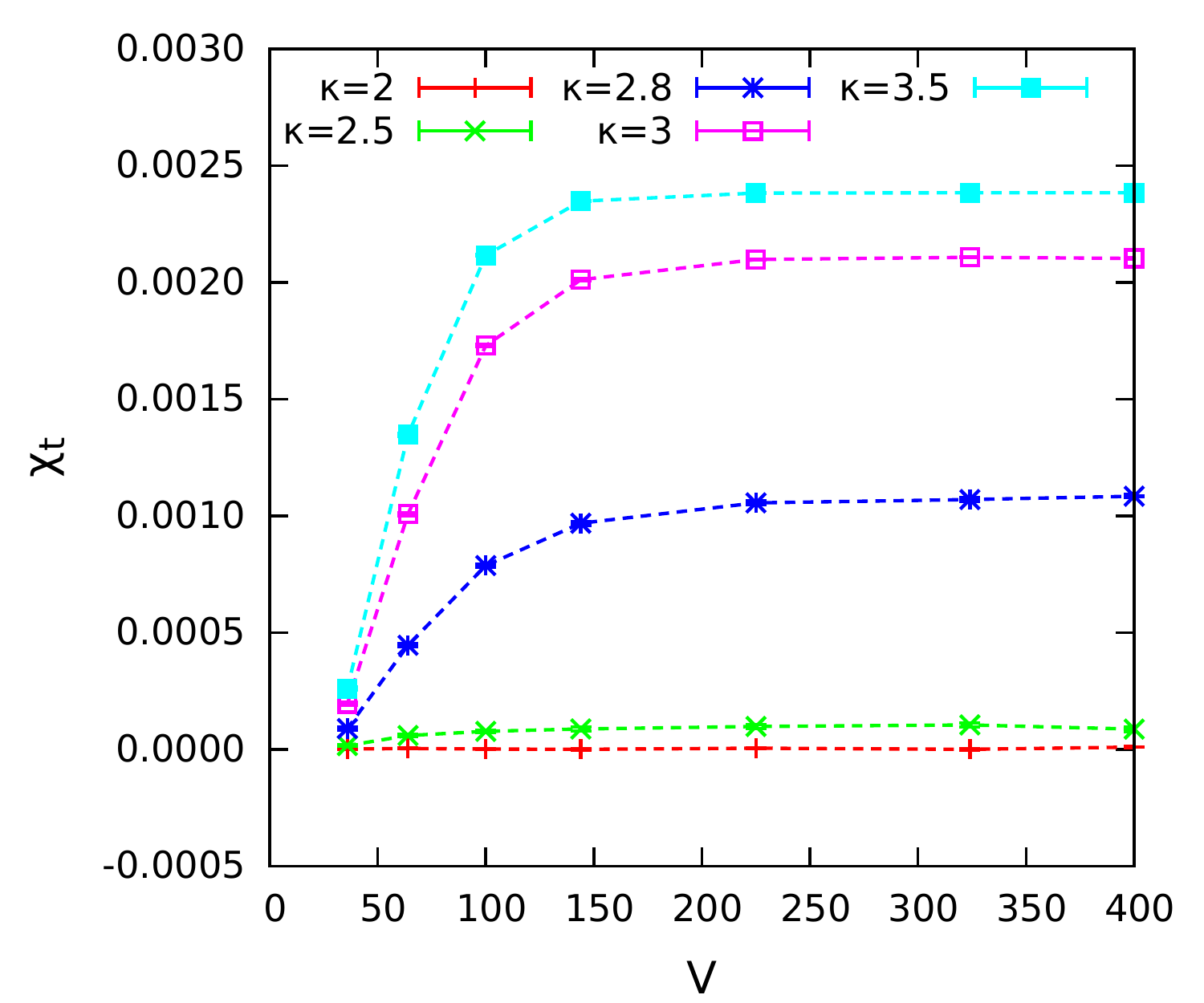}
\includegraphics[height=6cm,type=pdf,ext=.pdf,read=.pdf]{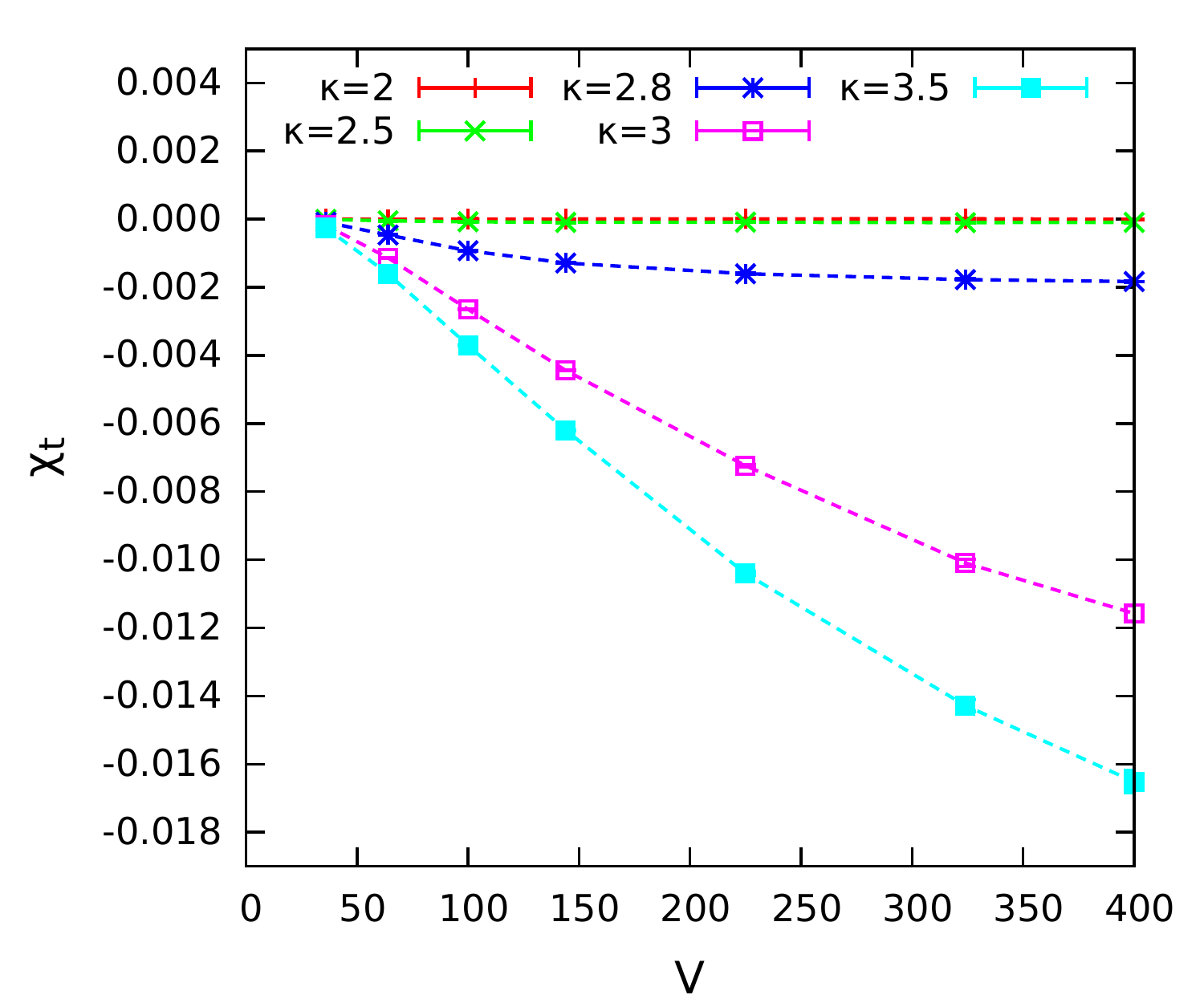}
\end{center}
\caption{Topological susceptibility versus the volume \(V=N_s\times N_t\) for different \(\kappa\) at \(\lambda=0.5\) and 
\(\beta=10\) at \(\theta=0\) (lhs) and \(\theta=\pi\) (rhs).}
\label{fig_chitop_vs_vol}
\end{figure*}

In Fig. \ref{fig_chitop_vs_vol} we now look at  the volume dependence of $\chi_{t}$ at \(\theta=0\) and \(\theta=\pi\) and compare 
the behavior for different values of the mass parameter $\kappa$.
The topological charge density $\langle q \rangle$ has a negative slope at \(\theta=0\) and therefore \(\chi_{t}\) is negative on the lhs.~plot. 
We observe a strong dependence on the mass parameter \(\kappa\). The susceptibility essentially vanishes in the broken phase 
and then starts to deviate from 0 for $\kappa \geq 2.5$ . For all values of \(\kappa\) a saturation is reached on lattice volumes between 
\(N_s=N_t=10\) and \(N_s=N_t=12\). For \(\theta=\pi\) (rhs.~plot) the behavior is different: Here $\chi_{t}$  also vanishes in the broken 
phase and then is positive for $\kappa \geq 2.5$. Most remarkably, $\chi_{t}$ does not seem 
to reach saturation as a function of the volume, a fact that hints at a possible phase transition. 

\begin{figure*}[p]
\begin{center}
\includegraphics[height=7cm,type=pdf,ext=.pdf,read=.pdf]{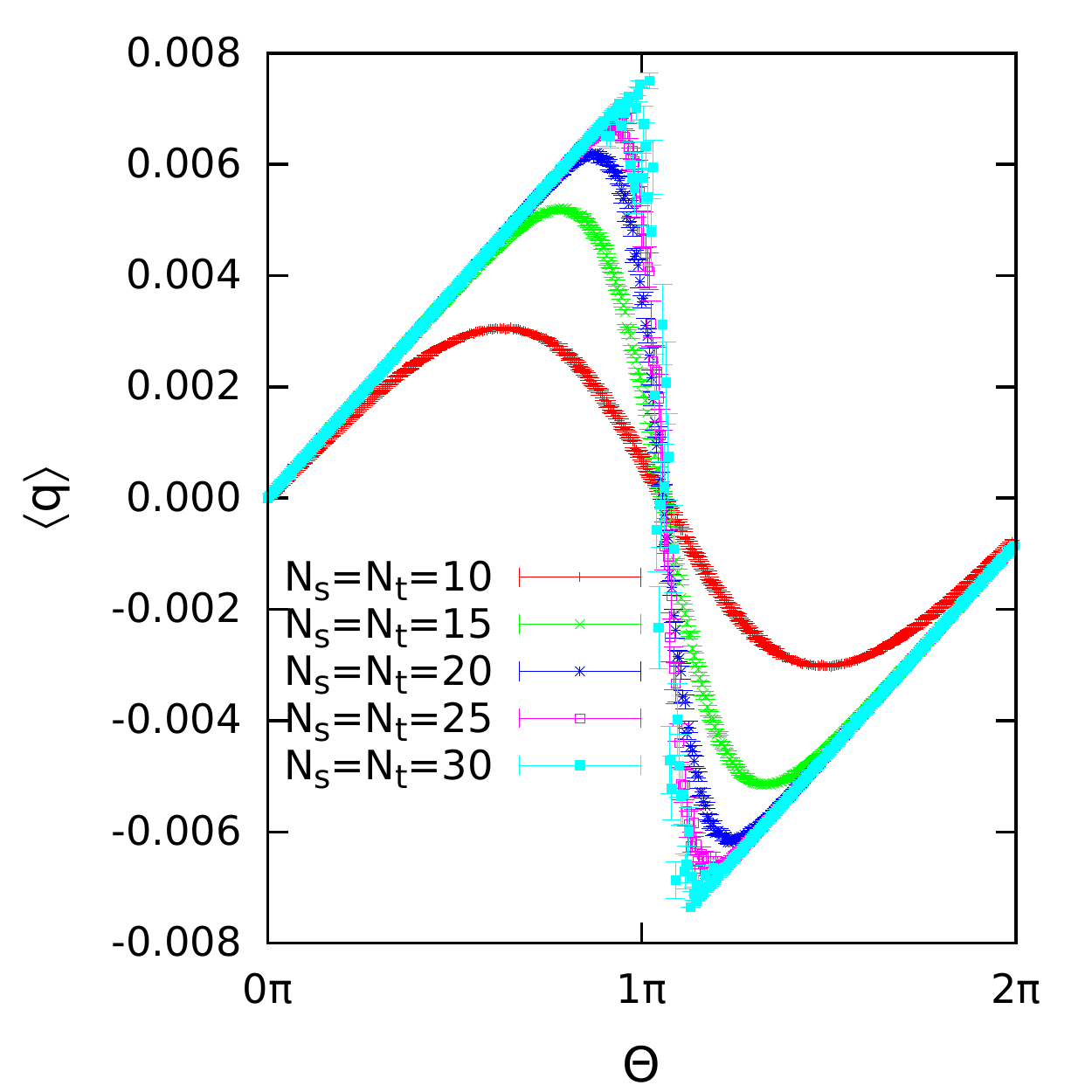}
\hskip8mm
\includegraphics[height=7cm,type=pdf,ext=.pdf,read=.pdf]{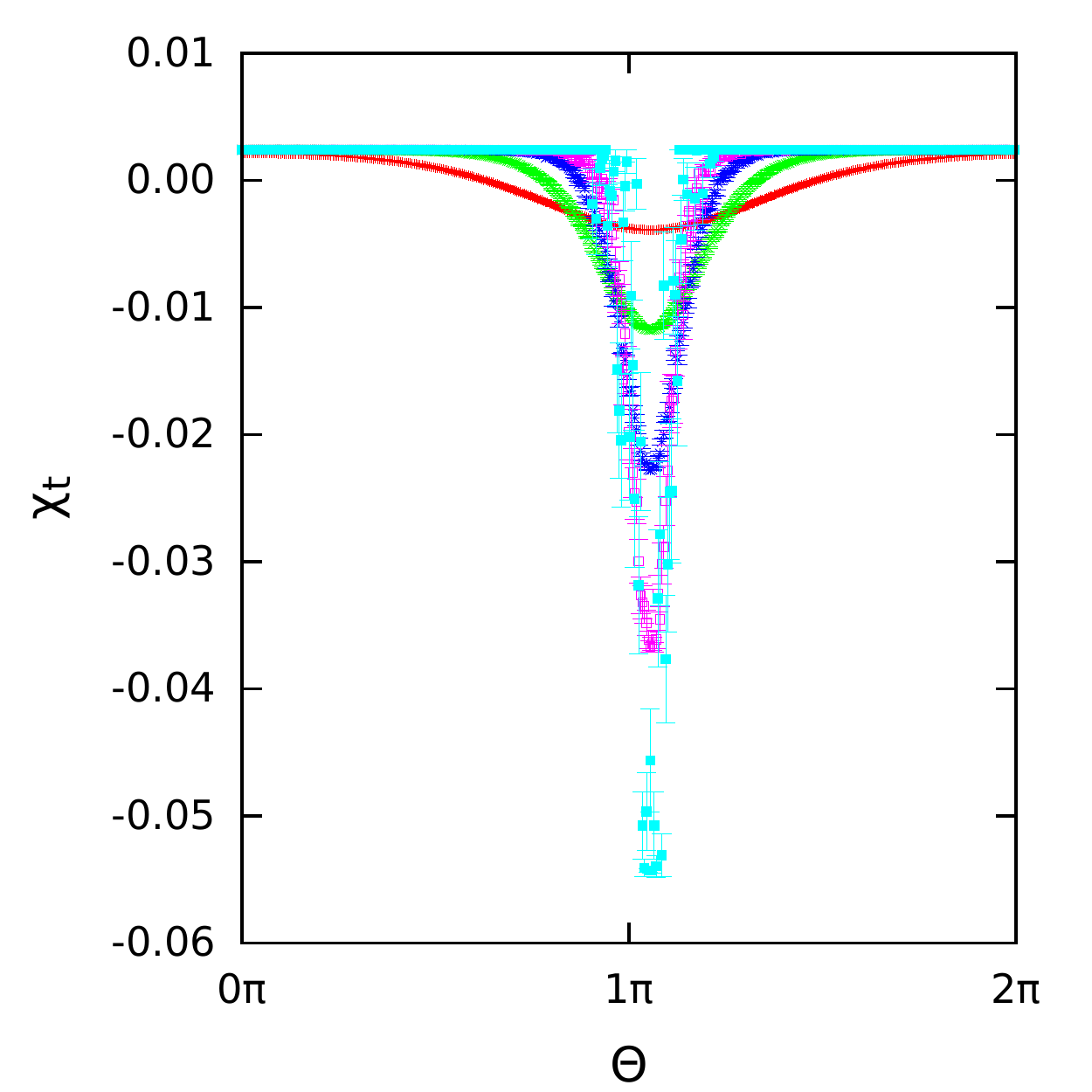}
\end{center}
\vspace{-3mm}
\caption{Topological charge and $\chi_{t}$  versus \(\theta\) at \(\kappa=4.0\), \(\lambda=0.5\) and \(\beta=10.0\) for different lattice volumes.}
\label{fig_top_chitop_vs_t_phase_transition}
\vskip5mm
\begin{center}
\includegraphics[height=5.8cm,type=pdf,ext=.pdf,read=.pdf]{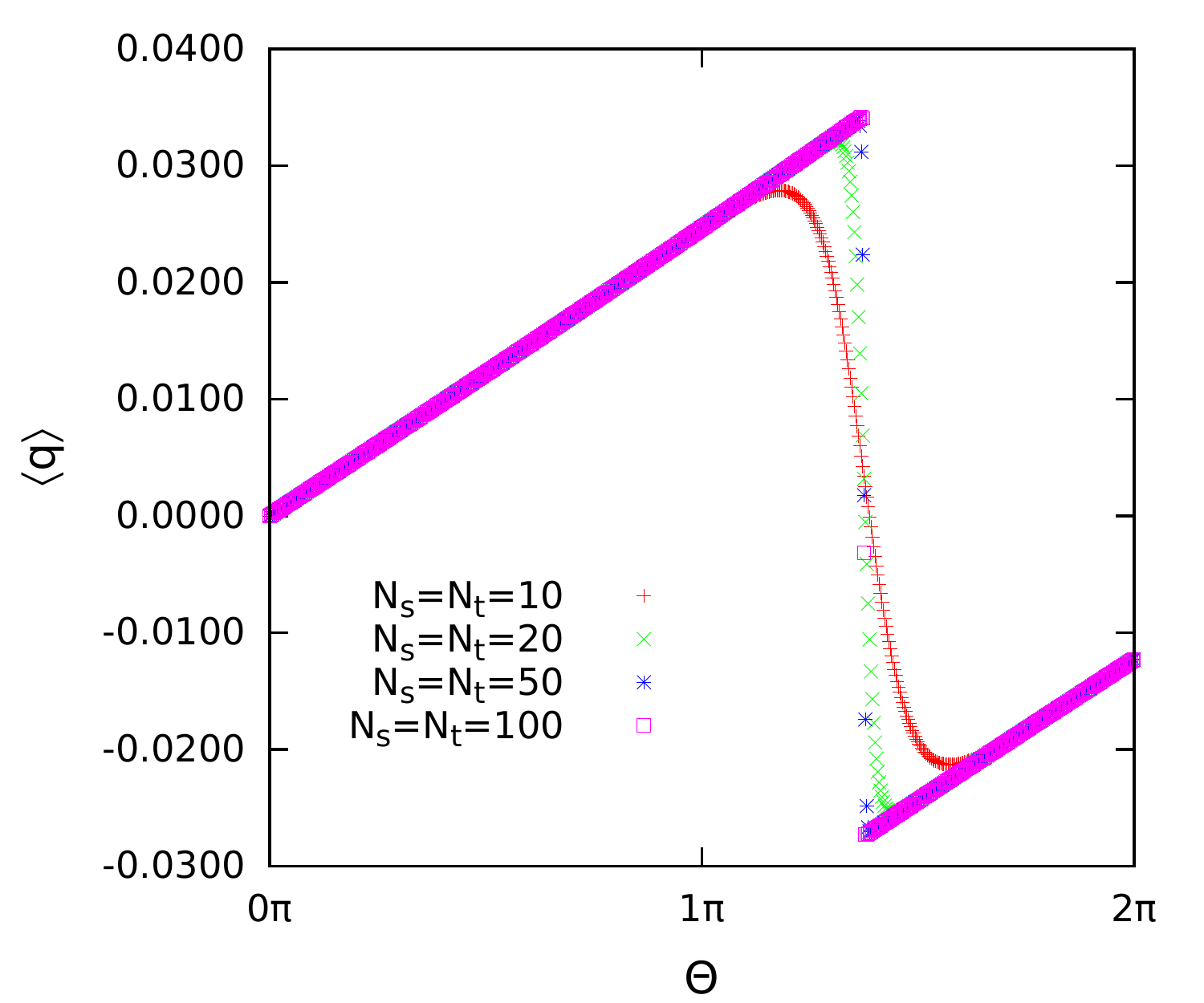}
\hskip8mm
\includegraphics[height=5.8cm,type=pdf,ext=.pdf,read=.pdf]{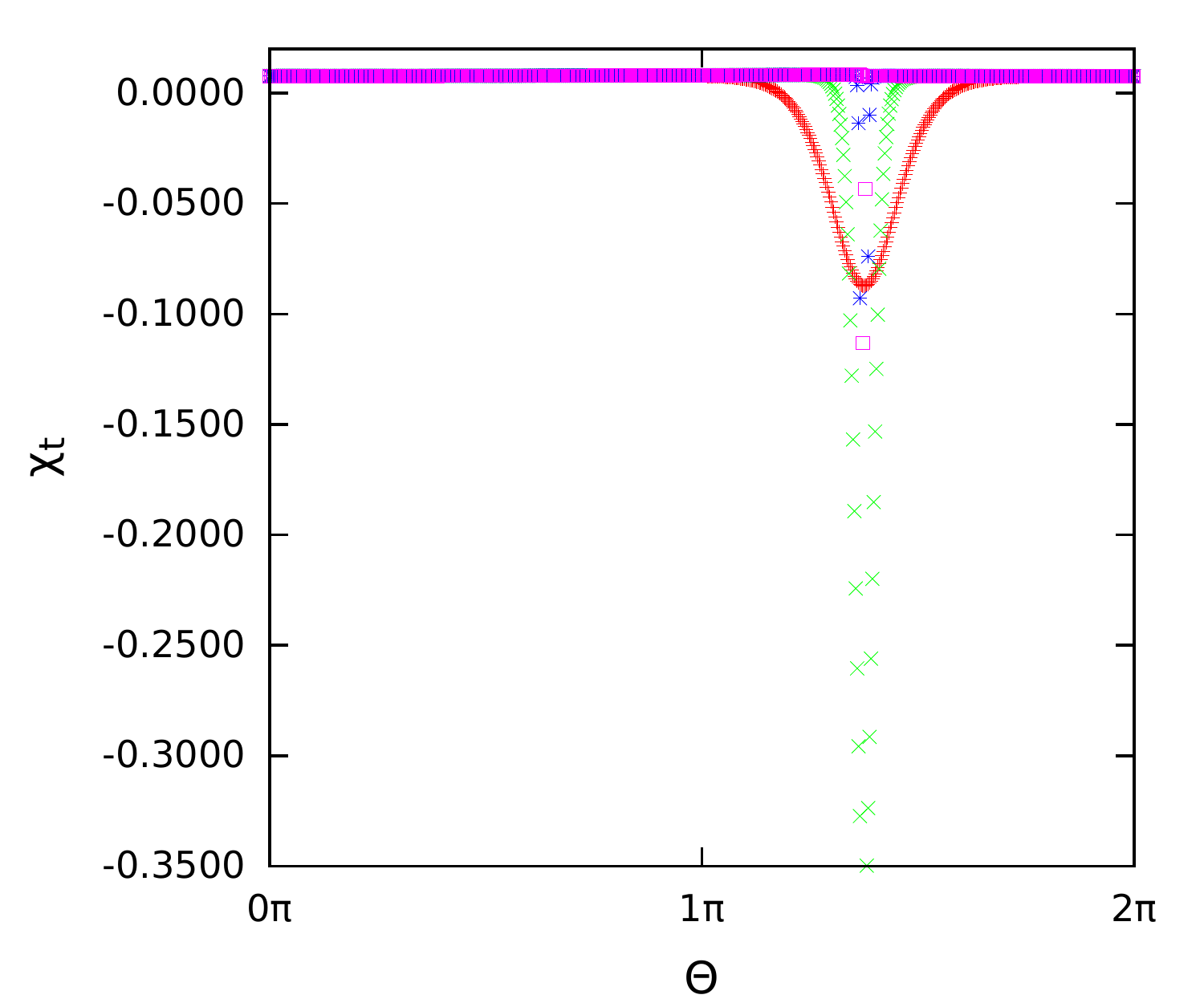}

\includegraphics[height=5.8cm,type=pdf,ext=.pdf,read=.pdf]{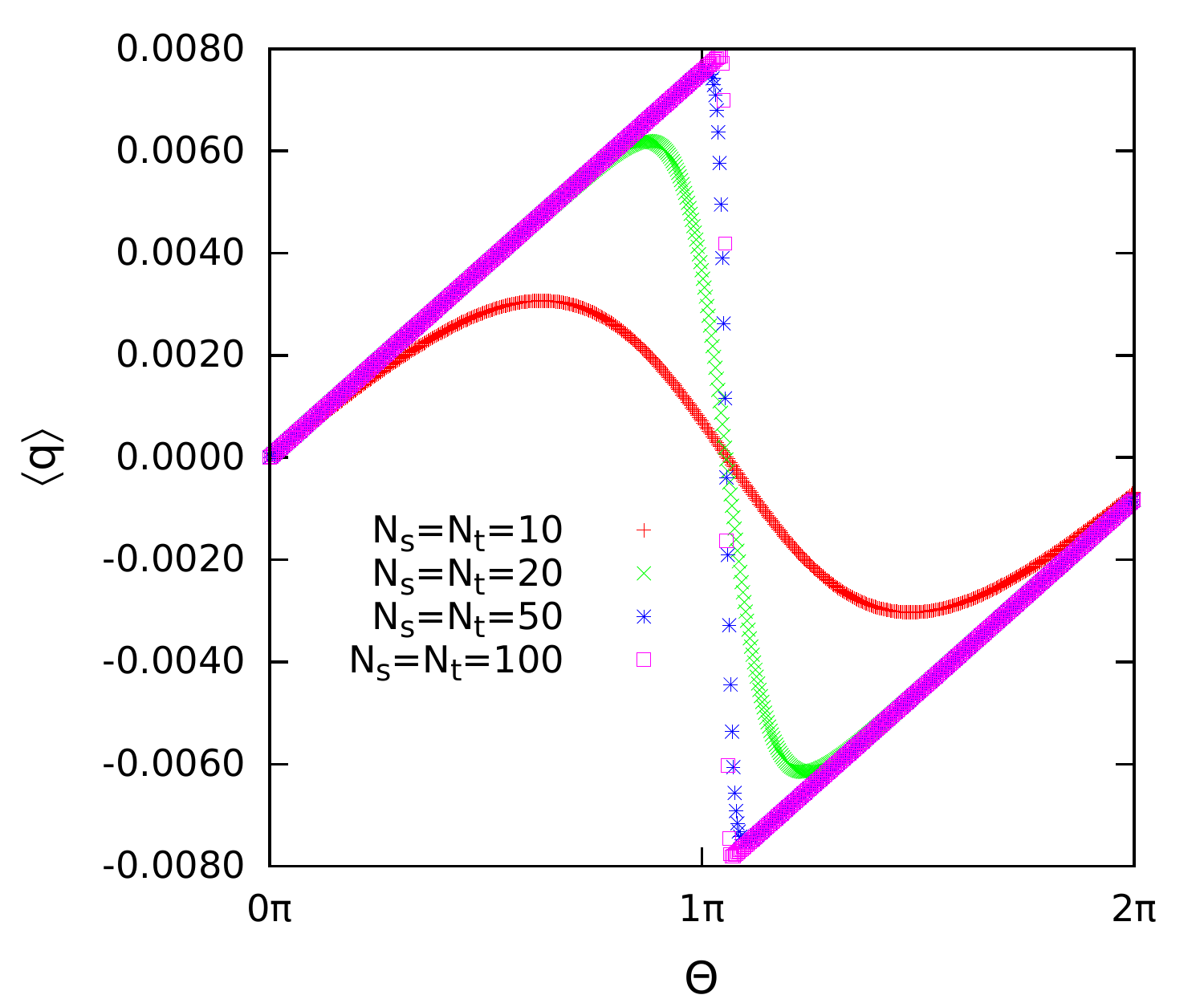}
\hskip8mm
\includegraphics[height=5.8cm,type=pdf,ext=.pdf,read=.pdf]{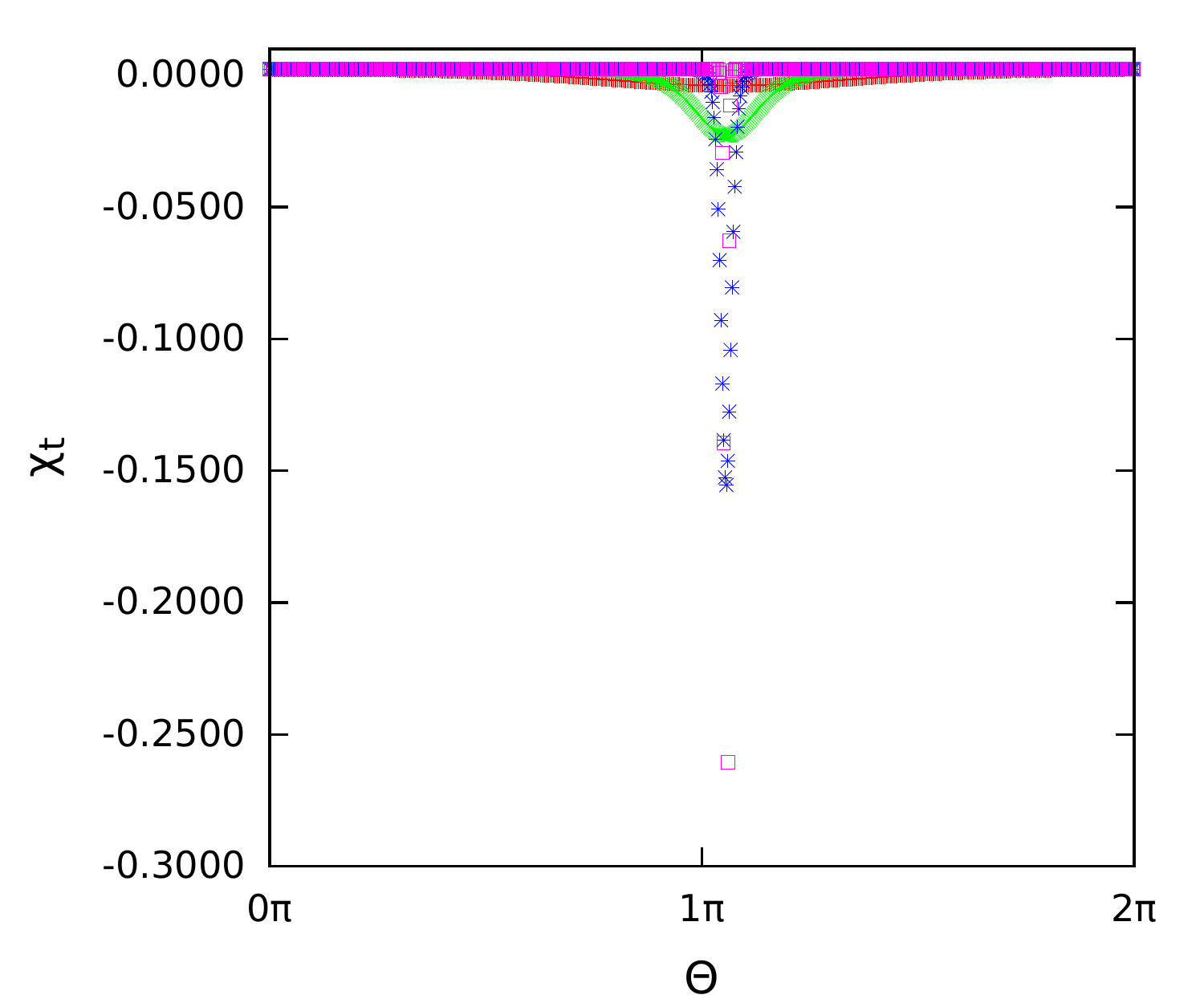}
\end{center}
\vspace{-3mm}
\caption{$\langle q \rangle$ (lhs.) and $\chi_{t}$ (rhs.) in pure gauge theory as a function of $\theta$.  
We show results at $\beta = 2.5$ (top) and $\beta = 10.0$ (bottom) and compare 
four different volumes.}
\label{puregauge_pi}

\end{figure*}

\subsection{The transition at $\theta = \pi$}
\label{sec_thetapi}

In the previous section we found evidence that in the symmetric phase there might be a transition at $\theta = \pi$. 
To identify the transition we analyzed the $\theta$-dependence of $\langle q \rangle$ and $\chi_{t}$ for a point in the symmetric phase,
in particular at $\lambda = 0.5$, $\kappa = 4.0$ with $\beta = 10.0$. The results as a function of $\theta$ for different volumes
are shown in Fig.~\ref{fig_top_chitop_vs_t_phase_transition}. It is obvious that $\chi_{t}$ (rhs.~plot) has a maximum near $\theta = \pi$,
and the height of the peak at the maximum increases with the volume. A detailed analysis shows, that the height of the maximum of 
$\chi_{t}$ scales almost perfectly with the volume, which indicates a first order transition. This is reflected in the behavior of 
$\langle q \rangle$, which in the large volume limit develops a discontinuity near $\theta = \pi$.
The analysis was repeated at other points in the symmetric phase with the same result and we conclude, 
that in the symmetric phase the system has a first order phase transition as a function of $\theta$.  

The transition at $\theta = \pi$ can be related to charge conjugation, i.e., the discrete symmetry transformation 
$U_{x,\mu} \rightarrow U_{x,\mu}^{\,*} \; \; \forall x, \mu$ and $\phi_x \rightarrow \phi_x^{\, *} \; \; \forall x$. While the action is invariant under this
symmetry, the topological charge changes sign: $Q[U] \rightarrow -Q[U]$. However, at $\theta = \pi$ the Boltzmann factor with the 
topological charge is given by $e^{-\.i\,\pi\, Q[U]} = (-1)^{Q[U]}$ which is again symmetric under $Q[U] \rightarrow - Q[U]$ and thus under charge 
conjugation. This discrete symmetry can be broken in the symmetric phase of the model. 

A careful inspection of Fig.~\ref{fig_top_chitop_vs_t_phase_transition} reveals that the transition is not located at exactly $\theta = \pi$, but 
appears at a slightly larger value of $\theta$. This can be attributed to an effect of finite lattice spacing.
To better understand this behavior we looked at the observables $\langle q \rangle$, $\chi_{t}$,
also in the pure gauge case where we can use the semi-analytical results 
from Section \ref{sec_puregauge}. The corresponding results for $\beta = 2.5$ and for $\beta = 10.0$ on four different volumes are shown in 
Fig.~\ref{puregauge_pi}. For the value $\beta = 10.0$ (plots in the bottom of Fig.~\ref{puregauge_pi}) we observe a behavior very similar to the
one of Fig.~\ref{fig_top_chitop_vs_t_phase_transition}. However, at $\beta = 2.5$, where we are further away from the continuum limit, the transition
is seen at an even larger value of $\theta$ and indicates that the position $\theta_{crit} = \pi$ for the transition 
is reached only in the continuum limit, i.e., for $\beta \rightarrow \infty$. 

Only in the continuum limit $Q[U]$ becomes restricted to integers, while away from the continuum limit $Q[U]$ has a distribution that is 
localized not exactly around integers, but has its maxima at values shifted to slightly smaller numbers than the integers of the continuum limit. 
For example in the charge 1 sector one could find the maximum at $Q_{max}(\beta) = 0.8$ instead of the continuum limit value 
$Q_{max}(\infty) = 1$.  The shift of the transition towards values of $\theta$ larger than $\pi$ then is explained by the condition that 
$\theta_{crit} \times Q_{max}(\beta) = \pi$, such that the symmetry $U_{x,\mu} \rightarrow U_{x,\mu}^{\, *}$ emerges. This gives rise to 
$\theta_{crit} = \pi/Q_{\max}(\beta)$, which explains $\theta_{crit} > \pi$ for $\beta < \infty$.

We remark, that we also looked at other observables in the pure gauge case, in particular 
$\langle \, \mbox{Re} \, U_{x,p} \, \rangle$ and $\chi_{p}$. Also there we find a first order behavior at $\theta = \pi$.

\section{Summary and discussion}

In this exploratory study we explore strategies for using dual variables in a simulation of a lattice field theory with a topological term. We 
show for the case of the U(1) gauge-Higgs system in two dimensions (scalar Schwinger model) that a dual formulation can be found where
the complex action problem of the conventional representation is overcome. The dual variables are loops for the matter fields and surfaces for
the gauge fields and the partition sum has only real and positive contributions. We show that in terms of the dual variables a Monte Carlo simulation
is possible at finite values of the vacuum angle $\theta$. This constitutes the first example of a theory with a vacuum angle where a
simulation could be performed with a complete solution of the corresponding complex action problem.

Using the dual approach we show that for the plaquette expectation value and for the topological charge $\langle Q \rangle$ 
the expected $2 \pi$-periodicity emerges in a properly implemented continuum limit. This was shown for
the semi-analytically tractable case of pure gauge theory, as well as for the full theory with matter fields. For the latter case we also found that the 
behavior of the observables as a function of $\theta$ is essentially independent of the mass parameter $\kappa$ in the symmetric phase. 
In the broken phase we find a quantitative dependence on the mass parameter but no qualitative one, i.e., observables acquire a constant 
shift on top of which they show exactly the same periodic behavior as in the unbroken phase. A clear distinction between the 
Higgs- and the symmetric phase appears for $\theta = \pi$, where we identify a first order behavior of observables in the symmetric phase,
which is absent in the Higgs phase.

In \cite{Damgaard:1987ec} the same model was studied, however in four dimensions with an external source. 
There it was found that the Higgs phase splits into two regions, discriminated by the kind of magnetic flux penetration. 
Therefore we covered a large range of mass parameter values \(\kappa \in [-50,4]\) for \(\lambda=0.5\) to search 
for a hint of a phase change in the broken phase. However, the only transition behavior we saw, was the crossover from the 
symmetric to the Higgs phase as discussed in Section~\ref{sec_phase_diagram}. 
The topological charge as well as the topological susceptibility stay constant (within error bars) throughout the broken phase.

\section*{Appendix}
\subsection*{Derivation of the dual representation}

In this appendix we provide a brief summary of the steps leading to the dual representation (\ref{z_dual}) of the partition sum. 
Initially the derivation as given for abelian scalar and gauge fields in four dimensions in \cite{Gattringer:2012df,Gattringer:2012ap,Mercado:2013ola,Mercado:2013yta}, but the adaption of the 
arguments to two dimensions and the topological term is straightforward.

In (\ref{latticeZ}) we have defined the partition sum $Z_M[U]$ 
of the matter fields in a background configuration of lattice gauge fields $U_{x,\mu}$. 
For $Z_M[U]$ the dual partition sum can be directly taken over from \cite{Gattringer:2012df,Gattringer:2012ap,Mercado:2013ola,Mercado:2013yta},
\begin{align}
Z_M[U] \; = \; &\sum_{\{l,\bar{l}\}} \Bigg[ \prod_{x,\mu} \frac{(U_{x,\mu})^{l_{x,\mu}}}{(|l_{x,\mu}|+\bar{l}_{x,\nu})!\;\bar{l}_{x,\mu}!)} \Bigg]
\Bigg[ \prod_x P(n_x) \Bigg] \nonumber \\
& \times ~ \Bigg[ \prod_x \delta\Big(\sum_\mu (l_{x,\mu} - l_{x-\hat{\mu},\mu}) \Big) \Bigg] ,
\label{ZMdual}
\end{align}
with $l_{x,\mu} \in \mathds{Z}$, $\bar l_{x,\mu} \in \mathds{N}_0$ and 
$P(n_x)$ and $n_x$ as defined in Eq.~(\ref{p_n_def}). In its dual form $Z_M[U]$ is a sum over loops of conserved $l$-flux and the 
loops are dressed with the link variables via the terms $(U_{x,\mu})^{l_{x,\mu}}$.

The full partition sum $Z$ is obtained by integrating $Z_M[U]$ over the gauge fields 
$U_{x,\mu}$ with the Boltzmann factor $e^{-S_G[U]}$. The dressed loops in (\ref{ZMdual}) provide the additional gauge field dependent factor 
$\prod_{x,\mu} (U_{x,\mu})^{l_{x,\mu}}$ such that we need to compute the integral
\begin{align}
Z_G[l] &= \int \! \mathcal{D}[U] \Bigg[ \prod_{x,\mu} (U_{x,\mu})^{l_{x,\mu}} \Bigg] e^{-S_G[U]-i \, \theta \, Q[U]} \\
& = 
\int \! \mathcal{D}[U] \Bigg[ \prod_{x,\mu} (U_{x,\mu})^{l_{x,\mu}} \Bigg] \Bigg[ \prod_x \, e^{\, \eta \; U_{x,p}} \,  e^{\, \overline{\eta} \; U_{x,p}^*} 
\Bigg] , \nonumber
\end{align}
with $\eta$ and $\overline{\eta}$ defined in (\ref{etadef}). The dualization of the gauge fields proceeds in a way equivalent to the matter fields: 
The Boltzmann next step is to expand the exponentials at each lattice site and for the two terms in the exponent separately
\begin{widetext}
\begin{eqnarray}
Z_G[l] &\!=\!& \int \! \mathcal{D}[U] \Bigg[ \prod_{x,\mu}(U_{x,\mu})^{l_x,\mu} \prod_x 
\sum_{n_x=0}^{\infty} \frac{\eta^{n_x}}{n_x!}
\Big(U_{x,1} U_{x+\hat{1},2} U_{x+\hat{2},1}^* U_{x,2}^*\Big)^{n_x}
\sum_{\bar{n}_x=0}^{\infty} \; \frac{\bar{\eta}^{\bar{n}_x}}{\bar{n}_x!} \;     
\Big(U_{x,1}^*U_{x+\hat{1},2}^* U_{x+\hat{2},1} U_{x,2} \Big)^{\bar{n}_x} \Bigg] \nonumber\\
 &\!\!\!=\!\!\!&  \Bigg[ \prod_x \sum_{n_x, \bar{n}_x =0}^{\infty}\Bigg] \Bigg[ \prod_x \frac{\eta^{n_x}}{n_x!} \frac{\bar{\eta}^{\bar{n}_x}}{\bar{n}_x!} \Bigg] 
 \int \! \mathcal{D}[U] \Bigg[ \prod_x (U_{x,1})^{\; n_x-\bar{n}_x-n_{x-\hat{2}}+\bar{n}_{x-\hat{2}}+l_{x,1}} \; 
 (U_{x,2})^{\;\bar{n}_x-n_x-\bar{n}_{x-\hat{1}}+n_{x-\hat{1}}+l_{x,2}} \Bigg] \nonumber\\
  &\!\!\!=\!\!\!& \sum_{\{n,\bar{n}\}}\! \Bigg[ \! \prod_x \frac{\eta^{n_x}}{n_x!} \frac{\overline{\eta}^{\bar{n}_x}}{\bar{n}_x!}  \Bigg] \Bigg[ \prod_x 
 \int_{-\pi}^\pi \!\!\! \frac{dA_{x,1}}{2\pi} e^{iA_{x,1}( \, n_x-\bar{n}_x-n_{x\!-\!\hat{2}}+\bar{n}_{x\!-\!\hat{2}}+l_{x,1})}
 \int_{-\pi}^\pi \!\!\! \frac{dA_{x,2}}{2\pi} e^{iA_{x,2}( \, \bar{n}_x-n_x-\bar{n}_{x\!-\!\hat{1}}+n_{x\!-\!\hat{1}}+l_{x,2})} 
 \nonumber \\
 &\!\!\!=\!\!\!& \sum_{\{n,\bar{n}\}}\! \Bigg[ \! \prod_x \frac{\eta^{n_x}}{n_x!} \frac{\overline{\eta}^{\bar{n}_x}}{\bar{n}_x!}  \Bigg] \Bigg[ \prod_x 
 \delta\Big(n_x-\bar{n}_x-n_{x-\hat{2}}+\bar{n}_{x-\hat{2}}+l_{x,1}\Big)   \delta\Big(\bar{n}_x-n_x-\bar{n}_{x-\hat{1}}+n_{x-\hat{1}}+l_{x,2}\Big)\Bigg].
 \end{eqnarray}
\end{widetext}
Upon defining new summation variables \(p_x \in \mathds{Z}\) and \(\bar{p}_x \in \mathds{N}_0\)
\begin{equation}
p_x \equiv n_x-\bar{n}_x~,~~ |p_x|+2\bar{p}_x \equiv n_x+\bar{n}_x~,
\end{equation}
we get the partition function for the gauge fields in its final form
\begin{align}
&Z_G[l] \; = \; \sum_{\{p,\bar{p}\}} 
\Bigg[ \prod_x \frac{\eta^{\, (|p_x|+p_x)/2+\bar{p}_x} \; \overline{\eta}^{\, (|p_x|-p_x)/2+\bar{p}_x}}{(|p_x|+\bar{p}_x)!\;\bar{p}_x!} \Bigg] \nonumber \\
& \times ~\Bigg[ \prod_x \delta\Big(p_x-p_{x-\hat{2}}+l_{x,1}\Big) \; \delta\Big(p_{x-\hat{1}}-p_x+l_{x,2}\Big) \Bigg] \, .
\end{align}
Combining this result with the matter part from (\ref{ZMdual}) completes the derivation of the dual representation and yields the 
expression of Eq. (\ref{z_dual}).

\newpage
\bibliographystyle{apsrev}

\end{document}